\pdfoutput=1
\RequirePackage{iftex}
\documentclass{ieeetj}
\usepackage{fix-cm} 

\usepackage[T1]{fontenc}
\usepackage{etoolbox}
\usepackage{algorithmic}
\usepackage{algorithm}
\usepackage{array}
\usepackage{textcomp}
\usepackage{stfloats}
\usepackage{url}
\usepackage{verbatim}
\usepackage{graphicx}
\usepackage{amsfonts}
\usepackage{amsmath}
\usepackage{amssymb}
\usepackage{amsthm}
\usepackage{newtxmath}
\newtheoremstyle{publication}{6pt}{6pt}{\normalfont}{10pt}{\itshape}{:}{.5em}{}
\theoremstyle{publication}

\newtheorem{theo}{Theorem}
\newtheorem{rem}{Remark}
\usepackage{color}
\usepackage{booktabs}
\usepackage{multirow}
\usepackage[nocompress]{cite}
\makeatletter
\let\MYcaption\@makecaption
\makeatother
\usepackage[justification=centering,labelfont=sf,textfont=sf]{subcaption}
\makeatletter
\let\@makecaption\MYcaption
\makeatother
\usepackage{setspace}
\usepackage{calc} 
\input{10_style/newcommands.sty}
\usepackage{hyperref}
\hypersetup{
    colorlinks = true, 
    urlcolor = black, 
    linkcolor = black, 
    citecolor = black, 
    anchorcolor = black, 
    pdfborder = {0 0 0}, 
    breaklinks = true,  
}
\graphicspath{{12_fig/}}

\usepackage{flushend}

\ifPDFTeX\ifnum\pdfoutput>0
\pdfpagewidth=\paperwidth\pdfpageheight=\paperheight
\fi\fi
\hypersetup{citecolor=blue,linkcolor=blue,urlcolor=black}
\makeatletter
\def\normalsize{\@setfontsize{\normalsize}{10pt}{12pt}}
\normalsize
\def\titlefont{\sffamily\fontsize{22}{24}\selectfont\bfseries\centering}
\def\authorfont{\sffamily\fontsize{10}{13.5}\selectfont\bfseries\centering}
\def\corfont{\rmfamily\fontsize{6.6}{7.5}\selectfont\centering}
\makeatother
\makeatletter
\def\@IEEEinterspaceMINratioM{0.10}
\def\@IEEEinterspaceMINratioB{0.20}
\makeatother
\makeatletter
\patchcmd{\@maketitle}{\vspace*{-10.5pt}}{\vspace*{-4.704pt}}{}{}
\patchcmd{\@maketitle}{\afffont\@authornote}{\rmfamily\fontsize{6.6}{8.5}\selectfont\leftskip3pc plus1fill\rightskip3pc plus1fill\@authornote}{}{}
\patchcmd{\@maketitle}{\vskip27.5pt}{\vskip29.5pt}{}{}
\patchcmd{\@maketitle}{\addvspace{33.5pt}}{\addvspace{44pt}}{}{}
\makeatother
\AtBeginDocument{%
\abovedisplayskip=10pt plus2pt minus1pt
\belowdisplayskip=10pt plus2pt minus1pt
\abovedisplayshortskip=5pt plus2pt minus1pt
\belowdisplayshortskip=10pt plus2pt minus1pt
}

\newcommand{\OJColumnBreak}{\par\vskip-\lastskip\penalty-10000\relax}
\makeatletter
\newcommand{\OJContent}[1]{\begingroup
\renewenvironment{table}[1][]{\par\begingroup\def\@captype{table}}{\par\endgroup}
\renewenvironment{table*}[1][]{\par\begingroup\def\@captype{table}}{\par\endgroup}
\renewenvironment{figure}[1][]{\par\begingroup\def\@captype{figure}}{\par\endgroup}
\renewenvironment{figure*}[1][]{\par\begingroup\def\@captype{figure}}{\par\endgroup}
#1\par\endgroup}
\makeatother
\AtBeginEnvironment{itemize}{\setlength{\itemsep}{0pt}\setlength{\parsep}{0pt}\setlength{\topsep}{3pt}}

\makeatletter
\def\OJRunningHead{\vbox to0pt{\vskip-13bp
\hbox to\textwidth{\sffamily\fontsize{6.6}{7}\selectfont
YAMASHITA ET AL.: SAMPLING METHOD FOR GENERALIZED GRAPH SIGNALS WITH PRE-SELECTED VERTICES\hfil}
\vskip6bp\hrule height.3pt\vss}}
\def\OJFootOdd{\hbox to\textwidth{\rffont\thepage\hfil}}
\def\OJFootEven{\hbox to\textwidth{\rffont\hfil\thepage}}
\def\ps@headings{\let\@oddhead\OJRunningHead\let\@evenhead\OJRunningHead
\let\@oddfoot\OJFootOdd\let\@evenfoot\OJFootEven}
\def\ps@plain{\def\@oddhead{\vbox to0pt{\vskip-13bp
\hbox to\textwidth{\sffamily\fontsize{6.6}{7}\selectfont
IEEE Open Journal of Signal Processing\hfil Author-prepared version}
\vskip6bp\hrule height.3pt\vss}}\let\@evenhead\@oddhead
\def\@oddfoot{\hbox to\textwidth{%
\rlap{\raisebox{8bp}[0pt][0pt]{\vbox{%
\hbox to\textwidth{\hfil\sffamily\fontsize{6}{7}\selectfont
IEEE Open Journal of Signal Processing, vol. 7, pp. 314--323, 2026.\hfil}
\nointerlineskip\vskip4.0224bp
\hbox to\textwidth{\hfil\sffamily\fontsize{6}{7}\selectfont
\textcopyright\ 2026 The Authors. Licensed under CC BY 4.0. \url{https://creativecommons.org/licenses/by/4.0/}\hfil}}}}%
\OJFootOdd}}\let\@evenfoot\@oddfoot}
\makeatother
\hypersetup{pdftitle={Sampling Method for Generalized Graph Signals With Pre-Selected Vertices via DC Optimization},pdfauthor={Keitaro Yamashita, Kazuki Naganuma, Shunsuke Ono},pdfsubject={Author-prepared version: IEEE Open Journal of Signal Processing, vol. 7, pp. 314-323, 2026. DOI: 10.1109/OJSP.2026.3664335}}

\AtBeginDocument{\let\OJOriginalBibitem\bibitem
\renewcommand{\bibitem}[1]{\ifstrequal{#1}{yamashita2024generalized}{\OJColumnBreak\interlinepenalty10000\clubpenalty10000\widowpenalty10000}{}\OJOriginalBibitem{#1}}}

\newcommand{\OJDepthColumnBreak}[1]{\par\vskip-\lastskip\kern-#1\penalty-10000\relax}

\def\BibTeX{{\rm B\kern-.05em{\sc i\kern-.025em b}\kern-.08em
    T\kern-.1667em\lower.7ex\hbox{E}\kern-.125emX}}
\AtBeginDocument{\definecolor{tmlcncolor}{cmyk}{0.93,0.59,0.15,0.02}\definecolor{NavyBlue}{RGB}{0,86,125}}

\def\OJlogo{\sffamily\scriptsize IEEE Open Journal of Signal Processing}
\def\seclogo{}

\def\authorrefmark#1{\ensuremath{^{\textbf{#1}}}}

\begin{document}
\receiveddate{18 September 2025}
\reviseddate{31 January 2026}
\accepteddate{5 February 2026}
\publisheddate{13 February 2026}
\currentdate{24 February 2026}
\doiinfo{OJSP.2026.3664335}
\editor{The review of this article was arranged by Associate Editor M. Karzand.}
\jvol{7}\pubyear{2026}

\markboth{Sampling Method for Generalized Graph Signals With Pre-Selected Vertices via DC Optimization}{K. Yamashita {et al.}}

\title{Sampling Method for Generalized Graph\\Signals With Pre-Selected Vertices via\\DC Optimization}

\author{\makebox[\textwidth][c]{\scalebox{0.92}[1]{KEITARO YAMASHITA\authorrefmark{1} (Graduate Student Member, IEEE), KAZUKI NAGANUMA\authorrefmark{2} (Member, IEEE),}}\\AND SHUNSUKE ONO\authorrefmark{1} (Senior Member, IEEE)}
\affil{Institute of Science Tokyo, Tokyo 152-8550, Japan}
\affil{Tokyo University of Agriculture and Technology, Tokyo 183-8538, Japan}
\corresp{CORRESPONDING AUTHOR: KEITARO YAMASHITA (email: yamashita.k.45d1@m.isct.ac.jp).}
\authornote{This work was supported in part by JST FOREST under Grant JPMJFR232M, in part by JST AdCORP under Grant JPMJKB2307, in part by JST CREST under Grant JPMJCR25Q5, in part by JST ACT-X under Grant JPMJAX23CJ, in part by JSPS KAKENHI under Grant 22H03610, Grant 22H00512, Grant 23H01415, Grant 23K17461, Grant 24K03119, Grant 24K22291, Grant 25H01296, and Grant 25K03136, and in part by the Grant-in-Aid for JSPS Fellows under Grant 25KJ0117.}

\begin{abstract}
This paper proposes a method for vertex-wise aggregation sampling of a broad class of graph\linebreak
 signals, designed to attain the best possible recovery based on the generalized sampling theory. This is\linebreak
 achieved by designing a sampling operator by an optimization problem, which is inherently non-convex,\linebreak
 as the best possible recovery imposes a rank constraint. An existing method for vertex-wise aggregation\linebreak
 sampling is able to control the number of active vertices but cannot incorporate prior knowledge of mandatory\linebreak
 or avoided vertices. To address these challenges, we formulate the operator design as a problem that\linebreak
 handles a constraint on the number of active vertices and prior knowledge on specific vertices for sampling,\linebreak
 mandatory inclusion or exclusion. We transformed this constrained problem into a difference-of-convex (DC)\linebreak
 optimization problem by using the nuclear norm and a DC penalty for vertex selection. To solve this, we\linebreak
 develop a convergent solver based on the general double-proximal gradient DC algorithm. The effectiveness\linebreak
 of our method is demonstrated through experiments on various graph signal models, including real-world\linebreak
 data, showing superior performance in the recovery accuracy compared to existing methods.
\end{abstract}

\begin{IEEEkeywords}
    Difference-of-convex optimization, generalized sampling theory, graph signal processing, graph signal sampling, vertex-wise aggregation sampling.
\end{IEEEkeywords}


\maketitle

\section{INTRODUCTION}
\label{sec:intro}

  Graph signal processing (GSP) has emerged as a powerful framework for analyzing data on irregular structures, extending classical signal processing theories to the graph setting~\cite{narang2012perfect, agaskar2013spectral, shuman2015multiscale, tanaka2020gensamp}.
  This paradigm has proven invaluable in numerous fields by providing tools to handle complex and diverse data.
  Various research topics are associated with GSP, including graph learning~\cite{dong2016learning,egilmez2018graph,liu2019graph}, graph restoration and signal recovery~\cite{ono2015total,onuki2016graph,berger2017graph,nagahama2022graph,yamagata2025robust}.
  Its applications span various domains, such as social network analysis, brain network studies, transportation modeling, image analysis, artificial intelligence, and wireless sensor systems~\cite{cheung2018graph,hasanzadeh2019piecewise,zare2020fusing,takemoto2022graph}.

  A key aspect of GSP is graph signal sampling.
  Sampling enables efficient data storage and transmission by reducing the dimension of data and can reduce costs and energy consumption by limiting the number of active sensors in physical networks~\cite{tanaka2020sampling}.
  Unlike its classical counterpart in time-series signal processing, graph signal sampling is more challenging as it lacks a regular, shift-invariant structure~\cite{chen2015discrete,tanaka2020sampling}.
  Initial research focused on extending the Shannon-Nyquist theorem to the graph domain, leading to sampling bandlimited graph signals~\cite{chen2015discrete, marques2015sampling, anis2016efficient, tsitsvero2016signals, valsesia2018sampling, tanaka2018spectral, puy2018random, bai2020fast, jayawant2022practical}.
  However, the bandlimited assumption is often too restrictive for applications with localized features or sharp transitions, such as meteorological data in mountainous areas, landscape image data, and water distribution network data~\cite{dong2016learning,cheung2018graph,li2025sensor}.

  To address this limitation, the \textit{generalized sampling theory}~\cite{elder2009beyond, eldar2015sampling} has been extended to the graph domain.
  This theory provides a unified framework for achieving the best possible recovery of beyond bandlimited signals based on signal priors.
  This extension enables us to handle sampling beyond bandlimited graph signals~\cite{chepuri2018graphsamp,tanaka2018spectral, tanaka2020sampling,tanaka2020gensamp,hara2020generalized, hara2021design,hara2022gsss,hara2023graph,yamashita2024generalized,yamashita2025generalizedgraph,yamashita2025controlling}.

  Within the generalized sampling framework, a key challenge is to design a sampling operator that ensures high-precision recovery for beyond bandlimited graph signals while keeping the benefits of sampling, such as data compression and sensor reduction.
  Early methods focused on two main strategies: \textit{vertex-wise sampling}, which selects a subset of vertices~\cite{hara2022gsss}, and \textit{aggregation sampling}, which forms samples from a linear combination of multiple vertices~\cite{hara2021design,yamashita2024generalized,yamashita2025generalizedgraph}. 
  While the vertex-wise sampling~\cite{hara2022gsss} is efficient, it is sensitive to noise due to its high dependence on the selected vertices.
  Aggregation sampling such as~\cite{yamashita2024generalized} offers greater robustness but requires data from all vertices, negating the benefit of sensor reduction.
  To balance these trade-offs, \textit{vertex-wise aggregation sampling} was introduced, aggregating data from some specific vertices for sampling~\cite{yamashita2025controlling}.

  Despite these advances, a significant practical research gap remains.
  While the existing vertex-wise aggregation sampling method~\cite{yamashita2025controlling} can control the quantity of active vertices, it cannot handle prior knowledge about specific vertices whether they must be used or avoided for sampling.
  This limitation is critical in real-world applications.
  In a sensor network, certain sensors are required due to their high demands or strategic location; for instance, weather observation is legally mandatory at some airports by an international convention.
  Whereas, some should be avoided due to their durability limitations or high operational costs.
  Furthermore, the method in~\cite{yamashita2025controlling} proposes a non-convex optimization problem whose algorithm is not theoretically guaranteed to be convergent.
  This lack of a convergence guarantee can lead to an unstable sampling operator design.

  These limitations lead to the research question: \textit{How can we design a sampling operator for vertex-wise aggregation sampling that achieves the best possible recovery for beyond bandlimited graph signals, allows us to control the number of active vertices with prior knowledge of must-use and avoided vertices, and is based on a convergent algorithm?}

  Our main contributions to address this question are:

  \vspace{-0.3\baselineskip}
  \begin{itemize}
    \itemsep=0pt \parsep=0pt
      \setlength{\leftskip}{-4mm}
      \setlength{\labelsep}{1mm}
      \item
      Formulate a sampling operator design problem that allows for the pre-selection of two distinct vertex sets (sets of "must-use" vertices and "avoided" vertices) while controlling the total number of active vertices.
      \item
      Reformulate this constrained problem into a difference-of-convex (DC) optimization problem\footnote{
        DC optimization has recently been used in signal processing~\cite{naganuma2024hyperspectral,sato2024enhancing}.
      }
      with relaxing the constraint to achieve the best possible recovery and the constraint to control the number of active vertices.
      \item
      Develop a solver based on the general double-proximal gradient DC (GDPGDC) algorithm, which guarantees its convergence to a critical point of the proposed problem.
  \end{itemize}
  \vspace{-0.3\baselineskip}

  This paper is organized as follows. 
  Section~\ref{sec:preliminaries} reviews the fundamentals of graph signal sampling and the GDPGDC algorithm. 
  Section~\ref{sec:prop} details the proposed method. 
  Section~\ref{sec:experiments} presents experimental results to demonstrate its effectiveness. 
  Section~\ref{sec:conclusion} concludes the paper.

\par\kern14bp
\section{PRELIMINARIES}
  \label{sec:preliminaries}
  This section establishes the background of our proposed method.
  We review the principles of the generalized sampling for the graph domain.
  Then, we introduce the GDPGDC algorithm, which is the optimization tool to solve our proposed problem.
  A list of notations is provided in Table~\ref{table:notation}.
  \begin{table}[t]
    \captionsetup{font=scriptsize} 
    \centering
    \caption{{Notations and Definitions.}}
    \vspace{-5mm}
    \label{table:notation}
    \renewcommand{\arraystretch}{1}
    \small
    \begin{tabular}[t]{c|c}
        \toprule
        Notations & Definitions \\
        \midrule
        $\set{X}$ & a set \\ \hline
        $\vect$ & a vector  \\ \hline
        $\vele, \left[\vect\right]_{i}$ & the $i$-th element of $\vect$  \\ \hline
        $\Ltwo{\vect}$ & the $\ell_2$ norm of $\vect$, $\Ltwo{\vect} := \sqrt{\sum_i \vele^2}$  \\ \hline
        $\mat$ & a matrix  \\ \hline
        $\vmat{i}, \left[\mat\right]_{(i)}$ & the $i$-th row vector of $\mat$  \\ \hline
        $\ele, \left[\mat\right]_{ij}$ & the $(i,j)$-th element of $\mat$  \\ \hline
        $\tran{\mat}$ & the transpose of $\mat$  \\ \hline
        $\inv{\mat}$ & the inverse of $\mat$  \\ \hline
        $\pinv{\mat}$ & the pseudo-inverse of $\mat$  \\ \hline
        $\sval{\mat}{i}$ & the $i$-th singular value of $\mat$  \\ \hline
        $\FN{\mat}$ & the Frobenius norm of $\mat$, $\FN{\mat} := \sqrt{\langle \mat, \mat \rangle}$  \\ \hline
        $\NN{\mat}$ & the nuclear norm of $\mat$, $\NN{\mat} := \sum_{i} \sval{\mat}{i}$  \\ 
        \bottomrule
    \end{tabular}
    \vspace*{-5mm}
\end{table}

 
  \par\kern-7bp
  \subsection{GENERALIZED SAMPLING OF GRAPH SIGNALS}\label{subsec:gs}

  Here, we provide a concise overview of the generalized sampling framework for graph signals~\cite{eldar2015sampling,tanaka2020gensamp,tanaka2020sampling}, which forms the foundation of our approach.
  We consider a weighted, undirected graph with $\numv$ vertices.

  An original graph signal, denoted by $\gsig \in \gset \subseteq \realN$, is transformed into a sampled signal $\gsampsig \in \realM$ (where $\nums \leq \numv$) by a sampling operator $\sampMT \in \realMN$. 
  Thus, the sampled signal is defined as $\gsampsig := \sampMT \gsig$.
  To recover the signal, the sampled signal $\gsampsig$ is first processed by a correction operator $\corM$ and then mapped back to the graph by a reconstruction operator $\recM$.
  The final recovered signal $\grecsig \in \gset$ is given by
  \begin{equation}
  \label{gsig_recovery}
    \grecsig = \recM \corM \gsampsig = \recM \corM \sampMT \gsig.
  \end{equation}

  The reconstruction operator $\recM$ can be fixed due to computational or hardware constraints. 
  In this paper, we focus on the unconstrained case for simplicity.
  The optimal operators $\corM$ and $\recM$ to achieve the best possible recovery can be designed by the signal priors.
  In what follows, we provide an overview of the signal priors and the corresponding design of $\corM$ and $\recM$ in the generalized sampling theory from prior works~\mbox{\cite{tanaka2020gensamp,hara2023graph}}.

  \OJDepthColumnBreak{2.156bp}
  \noindent\begin{minipage}[t][661.8bp][t]{\columnwidth}
  \vspace*{-6.6bp}
  \abovedisplayskip=8.5pt \belowdisplayskip=8.5pt
  \abovedisplayshortskip=4pt \belowdisplayshortskip=8.5pt
  \subsubsection{SUBSPACE PRIOR}
    Under the subspace prior, we suppose that a graph signal $\gsig \in \realN$ is characterized by a linear model as follows:
    \begin{equation} \label{gs_sb}
        \gsig := \sbgenM \sbexpd,
    \end{equation}
    where $\sbgenM \in \realNK$ $(\sbgenn \leq \numv)$ is a known generator matrix and $\sbexpd \in \realK$ is an expansion coefficient vector. 
    Under this prior, with a given $\sampM$, it is known that $\grecsig$ is given as follows:
    \begin{equation}\label{sb_recsig}
        \grecsig = \sbgenM \pinv{(\sampMT \sbgenM)} \sampMT \gsig.
    \end{equation}
    Then, the operators $\corM$ and $\recM$ are given by
    \begin{equation} \label{sb_corM_recM}
        \corM = \pinv{(\sampMT \sbgenM)}, \quad \recM = \sbgenM.
    \end{equation}

  \subsubsection{SMOOTHNESS PRIOR}
    Under the smoothness prior, the signal $\gsig$ is assumed to be smooth as: $\Ltwo{\smopeM \gsig}^{2} \leq \smbound^2$ for some constant $\smbound$, where $\smopeM \in \realNN$ is the invertible operator that quantifies the variation of $\gsig$.
    Under this prior, with a given $\sampM$, it is known that $\grecsig$ is given as follows:
    \begin{equation}\label{sm_recsig}
      \grecsig = \smMXrecM \pinv{(\sampMT\smMXrecM)} \sampMT \gsig, 
    \end{equation}
    where $\smMXrecM = (\smopeMT\smopeM)^{-1}\sampM$.
    Then, the operators $\corM$ and $\recM$ are given by
    \begin{equation} \label{sm_corM_recM}
        \corM = \pinv{(\sampMT \smMXrecM)}, \quad
        \recM = \smMXrecM = \inv{(\smopeMT\smopeM)}\sampM.
    \end{equation}

  \subsubsection{STOCHASTIC PRIOR}
    Under the stochastic prior, we consider that the samples can be obtained with an additive noise $\stnoise \in \realM$ and recover from the noisy sampled signal $\ssign = \gsampsig + \stnoise$ as follows: 
    \begin{equation} \label{st_reconstruction}
      \grecsig = \recM \corM \ssign = \recM \corM (\gsampsig + \stnoise) = \recM \corM (\sampMT \gsig + \stnoise).
    \end{equation}

    Suppose a graph signal $\gsig$ as a zero-mean process characterized by graph wide sense stationarity (GWSS)~\cite{hara2023graph}.
    The graph signal $\gsig$ and the noise $\stnoise$ is associated with a known covariance matrix $\stsigM \in \realNN$ and $\stnoiM \in \realMM$, respectively.
    Both $\stsigM$ and $\stnoiM$ are autocorrelation matrices.
    Under this prior, with a given $\sampM$, it is known that $\grecsig$ is given as follows:
    \begin{equation}\label{st_recsig}
      \grecsig = \stsigM \sampM \pinv{(\sampMT \stsigM \sampM + \stnoiM)} (\sampMT \gsig + \stnoise).
    \end{equation}
    Then, the operators $\corM$ and $\recM$ are given by
    \begin{equation} \label{st_corM_recM}
      \corM = \pinv{(\sampMT \stsigM \sampM + \stnoiM)}, \quad
      \recM = \stsigM \sampM.
    \end{equation}

  \vspace{-5bp}
  \subsection{GENERAL DOUBLE-PROXIMAL GRADIENT DC ALGORITHM}

  This subsection details the \textit{general double-proximal gradient DC} (GDPGDC) algorithm~\cite{banert2019general}, a versatile and convergent method for solving optimization problems involving a DC structure.
  The GDPGDC algorithm is designed to address problems of the form
  \abovedisplayskip=6pt \belowdisplayskip=2pt
  \begin{equation}
    \label{prob:DC_general}
    \min_{\mat} f_1(\mat) + f_2(\mat) - h(\matDC \mat),
  \end{equation}
  \par\kern-3.111bp
  where 
    $f_1: \realnn{n}{m} \rightarrow \realnum$ is a differentiable convex function with a $1/\beta$-Lipschitz continuous gradient for some $\beta >$
  \par\end{minipage}\OJColumnBreak
  \noindent\begin{minipage}[t][67bp][t]{\columnwidth}\vspace*{-6.17bp}\OJContent{\begin{table}[t]
    \captionsetup{font=scriptsize}
    \centering
    \caption{{Definitions of the Matrix $\varM$ for Each Prior. The Following Definitions Are Used: $\smopeM = \smsvul \smsv \smsvur^\top$ and $\stsigM = \stsqM^\top \stsqM$}.}
    \vspace{-5mm}      
    \label{tab:summary_varM}
    \renewcommand{\arraystretch}{1}
    \footnotesize
    \begin{tabular}[t]{c|ccc}
        \toprule
             &Subspace&Smoothness&Stochastic\\ \hline
            $\varM$ & $\sbgenMT$  & $\smsv^{-1} \smsvur^{\top}$ & $\stsqM$ \\
        \bottomrule
    \end{tabular}
    \vspace{-3mm}
\end{table}}\end{minipage}\par\noindent
  $0$, 
    $f_2:\realnn{n}{m} \rightarrow \exR$ and 
    $h:\realnn{k}{m} \rightarrow \exR$ are 
      proper, lower-semicontinuous, convex functions,
      with $\exR := \realnum \cup \{\infty\}$, and 
    $\matDC: \realnn{k}{n}$ is a linear operator.

  The GDPGDC algorithm solves Prob.~\eqref{prob:DC_general} through an iterative process.
  Given stepsize parameters $\gamma_1^{(t)}, \gamma_2^{(t)} > 0$ and starting from an initial point $(\mat^{(0)}, \DCw^{(0)})$, the algorithm proceeds by alternating between updates for the primal variable $\mat$ and a dual variable $\DCw$:
  \begin{equation}
      \label{eq:DCAbyGDPG}
      \resizebox{\linewidth-0mm}{!}{$
      \left\lfloor
      \begin{array}{l}
          \mat^{(\Step + 1)} \leftarrow
            \prox_{\gamma_{1}^{(\Step)} f_{2}} \left(
              \mat^{(\Step)} - \gamma_{1}^{(\Step)} \left(\nabla f_{1} \left(\mat^{(\Step)} \right) - \matDCT \DCw^{(\Step)} \right)
            \right); \\
          \DCw^{(\Step+1)} \leftarrow
            \prox_{\gamma_{2}^{(\Step)}h^{*}} \left(
              \DCw^{(\Step)} + \gamma_{2}^{(\Step)}\matDC \mat^{(\Step + 1)}
            \right); \\
          \Step \leftarrow \Step + 1.
      \end{array}
      \right.
      $}
  \end{equation}

  Here, the proximity operator $\prox_{\gamma f}(\cdot)$ for a proper, lower-semicontinuous, convex function $f$ and a parameter $\gamma > 0$ is defined as
  \begin{equation} \label{eq:prox_f2}
    \prox_{\gamma f}(\mat) := \argmin_{\matt} \left\{ f(\matt) + \frac{1}{2\gamma} \FN{\mat - \matt}^{2} \right\}.
  \end{equation}

  The update for $\DCw$ involves the proximity operator of $h^*$, the 
  \textit{Fenchel–Rockafellar conjugate function}\footnote{
    The \textit{Fenchel–Rockafellar conjugate function} of $h$ is defined as \\ $h^*(\mat) :=\max_{\matt}\langle \mat, \matt \rangle - h(\matt)$, where $\langle \mat, \matt \rangle$ denotes the inner product of $\mat$ and $\matt$, $\langle \mat, \matt \rangle := \sum_{i}\sum_{j} \ele \elee$.
  } of $h$.
  This can be computed by using Moreau's identity~\cite[Theorem~14.3(ii)]{bauschke2017correction} as
  \begin{equation}
    \label{eqDCh}
    \prox_{\gamma h^*}(\mat) = \mat - \gamma \prox_{\frac{1}{\gamma} h}\left(\tfrac{1}{\gamma}\mat\right).
  \end{equation}

  The convergence properties of this algorithm are well-established. 
  The following theorem provides the conditions under which the sequence generated by the GDPGDC algorithm converges to a critical point.
  \begin{theo}[{\cite[Proposition~4]{banert2019general}} Convergence of GDPGDC]
  \label{theo:convergence_GDPGDC}
    Assume that the objective function in Prob.~\eqref{prob:DC_general} is bounded from below. 
    Let the step-sizes satisfy $0 < \gamma_1^{(\Step)} < 2 \beta$ and $0 < \gamma_2^{(\Step)} < +\infty$. 
    If the sequences $\{\mat^{(\Step)}\}_{\Step \in \mathbb{N}}$ and $\{\mathbf{Z}^{(\Step)}\}_{\Step \in \mathbb{N}}$ generated by the iterative scheme~\eqref{eq:DCAbyGDPG} are bounded, then any cluster point of $\{\mat^{(\Step)}\}_{\Step \in \mathbb{N}}$ is a critical point of Prob.~\eqref{prob:DC_general}.
  \end{theo}

\section{PROPOSED METHOD}
\label{sec:prop}

  In this section, we present our method for designing a sampling operator $\sampM$, which selects a subset of the vertices by its row-sparsity and aggregates values from the selected vertices.
  The sampling and recovery process in \eqref{gsig_recovery} relies on a correction operator $\corM$ and a reconstruction operator $\recM$, which are designed for each signal prior as introduced in Section~\ref{sec:preliminaries}.\ref{subsec:gs}.
  Our goal is to design a $\sampM$ that is appropriate for the process with accommodating prior knowledge about which vertices are must-use or avoided for sampling.

  \subsection{THE CONDITION FOR DESIGNING THE SAMPLING OPERATOR}
  \label{subsec:condition_for_sampling_operator}
  In this subsection, we discuss the condition for designing a sampling operator on prior-by-prior basis by using the solutions about recovered signals introduced in Section~\ref{sec:preliminaries}.\ref{subsec:gs}. before discussing the specific designing method.
  In order to achieve the best possible recovery based on Eqs.~\eqref{gsig_recovery} and~\eqref{st_reconstruction}, and the recovery process to be numerically stable and for the resulting optimal solution to be unique, the matrices involved in the pseudo-inverse operation in Eqs.~\eqref{sb_recsig},~\eqref{sm_recsig}, and~\eqref{st_recsig} should have the full rank.
  We will now examine the conditions on $\sampM$ for each signal prior to ensure that these matrices have the full rank.

  \subsubsection{SUBSPACE PRIOR} \label{prop:sb}
  Based on~\eqref{sb_recsig}, we seek $\sampM$ for which $\pinv{(\sampMT\sbgenM)}$ has the full row rank\footnote{For simplicity, we assume $\nums \geq \sbgenn$.}, i.e., $\sampMT \sbgenM$ has the full column rank.
  Thus, we need to find $\sampM$ such that $\sbgenMT \sampM$ has the full row rank.

  \subsubsection{SMOOTHNESS PRIOR} \label{prop:sm}
  Based on~\eqref{sm_recsig}, we seek $\sampM$ for which $\pinv{(\sampMT \smMXrecM)} = \inv{(\sampMT \smMXrecM)}$, i.e., $\sampMT \smMXrecM$ is invertible.
  Let us define the singular value decomposition (SVD) of $\smopeM$ as $\smopeM = \smsvul \smsv \smsvur^\top$.
  Then, $\sampMT \smMXrecM$ is transformed as follows:
  \begin{align}
      \sampMT \smMXrecM &= \sampMT (\smopeMT\smopeM)^{-1}\sampM 
      = \sampMT (\smsvur \smsv^{2} \smsvur^\top)^{-1}\sampM \nonumber \\
      &= (\smsv^{-1} \smsvur^\top \sampM)^\top \smsv^{-1} \smsvur^\top \sampM.
  \end{align}
  Thus, we need to find $\sampM$ such that $(\smsv^{-1} \smsvur^\top \sampM)^\top \smsv^{-1} \smsvur^\top \sampM$ is invertible, i.e., $\smsv^{-1} \smsvur^\top \sampM$ has the full rank.

  \subsubsection{STOCHASTIC PRIOR} \label{prop:st}
  Based on~\eqref{st_recsig}, we seek $\sampM$ for which $\pinv{(\sampMT\stsigM\sampM + \stnoiM)} = \inv{(\sampMT\stsigM\sampM + \stnoiM)}$, i.e., $\sampMT\stsigM\sampM + \stnoiM$ is invertible.
  Due to the characteristics of an autocorrelation matrix, $\sampMT\stsigM\sampM + \stnoiM$ is invertible if $\sampMT\stsigM\sampM$ is invertible.
  Since $\stsigM$ is symmetric, $\stsigM$ can be decomposed as $\stsigM = \stsqM^\top \stsqM$ for some $\stsqM \in \realNN$.
  Thus, we need to find $\sampM$ such that $\sampMT\stsigM\sampM = (\stsqM \sampM)^\top \stsqM \sampM$ is invertible, i.e., $\stsqM \sampM$ has the full rank.

  \medskip
  Based on the above discussions, we introduce a matrix $\varM$ to find an appropriate $\sampM$ to unify the design conditions for all signal priors.
  To unify the conditions introduced in~\ref{prop:sb}),~\ref{prop:sm}), and~\ref{prop:st}), we seek $\sampM$ such that $\varM \sampM$ has the full rank, where the definition of $\varM$ depends on the signal prior as summarized in Table~\ref{tab:summary_varM}.

  \subsection{PROBLEM FORMULATION}

  To incorporate such prior knowledge about the vertices, we partition the set of all vertices $\gver$ into three disjoint sets: the set of must-use vertices, $\setSamp$; the set of avoided vertices, $\setNsamp$; and the set of vertices which are unknown whether they are to be active, $\setUsamp$, whose vertices are the remaining vertices not included in $\setSamp$ or $\setNsamp$.
  We also define $\numz$ $(\numz \leq \numv)$ as the desired upper limit on the total number of active vertices.
  From this, the upper limit on the number of active vertices to be selected from $\setUsamp$ is denoted by $\numAdd = \numz - |\setSamp|$.

  To achieve the best possible signal recovery, $\sampM$ should satisfy the ideal conditions derived from the generalized sampling theory; 
  in particular, $\varM \sampM$ has the full rank.
  Additionally, we must enforce the structural constraints on $\sampM$ based on our vertex partitioning.

  First, given $\varM$, $\setSamp$, $\setNsamp$, and $\numz$, we formulate the problem of finding a sampling operator $\sampM$ that satisfies the ideal conditions as a feasibility problem as follows:
  \begin{align} \label{prob:org}
      \text{find } \sampM \quad \text{s.t.} \quad
      \begin{cases}
        \sampM \in \setNsampM, \\
        \sampM \in \setUsampM, \\
        \rank(\varM \sampM) = R,
      \end{cases}
  \end{align}
  where $\setNsampM := \{ \mat \mid \vmat{i} = \zeros, \forall i \in \setNsamp \}$ is the set of matrices respecting the avoided vertices constraint, $\setUsampM := \{ \mat \mid | \{ i \in \setUsamp \mid \Ltwo{\vmat{i}} \neq 0 \} | \leq \numAdd \}$ restricts the number of active vertices from the undecided set $\setUsamp$, and $R$ is the largest possible rank of $\varM \sampM$, where $R = \sbgenn$ for the subspace prior and $R = \nums$ for the smoothness and stochastic priors.

  However, the rank constraint and the combinatorial vertex selection constraint in~\eqref{prob:org} are intractable to handle directly.
  As the first step, we relax the hard rank constraint into the tightest convex approximation of the rank function that we use the nuclear norm and reframe the problem as a constrained maximization by following~\cite{fazel2002matrix}:
  \begin{align} \label{prob:org2}
    \max_{\sampM} \NN{\varM \sampM} \quad \text{s.t.} \quad
    \begin{cases}
      \sampM \in \setNsampM, \\
      \sampM \in \setUsampM.
    \end{cases}
  \end{align}
  This \eqref{prob:org2} is equivalent to the following minimization problem by introducing the indicator functions\footnote{The indicator function $\iota_{\set{C}}(\mat)$ for a closed set $\set{C}$ is defined as $\iota_{\set{C}}(\mat) = 0$ if $\mat \in \set{C}$ and $\iota_{\set{C}}(\mat) = +\infty$ otherwise.}
   for the sets $\setNsampM$ and $\setUsampM$:
  \begin{align} \label{prob:org3}
    \min_{\sampM} \iota_{\setNsampM}(\sampM) + \iota_{\setUsampM}(\sampM) - \NN{\varM \sampM}.
  \end{align}

  Next, we address the non-convex combinatorial constraint $\iota_{\setUsampM}(\sampM)$.
  We replace this intractable indicator function with a DC penalty term that promotes sparsity among the rows corresponding to the set $\setUsamp$.
  Then, \eqref{prob:org3} is transformed into the following formulation:
  \begin{align} \label{prob:org4}
    \min_{\sampM} \iota_{\setNsampM}(\sampM) + \lambda \left( \sum_{i \in \setUsamp} \Ltwo{ \sampvec_{(i)}} - \Omega_{\numAdd} (\sampM_{\setUsamp})\right) - \NN{\varM \sampM},
  \end{align}
  where $\sampM_{\setUsamp}$ is the sub-matrix of $\sampM$ with rows corresponding to $\setUsamp$, $\Omega_{\numAdd}(\mat)$ is a function summing the $\ell_2$ norms of the $\numAdd$ largest rows of $\mat$, and $\lambda > 0$ is a parameter.

  Finally, we add a regularization term $\frac{\delta}{2} \FN{\sampM}^{2}$ ($\delta > 0$) to ensure the objective function is bounded below for avoiding $\sampM$ to be arbitrarily large.
  By rearranging the terms, our final DC optimization problem is formulated as:
  \begin{align} \label{prob:prop}
    \begin{split}
      \min_{\sampM}
      & \left( \iota_{\setNsampM}(\sampM)
        + \lambda \sum_{i \in \setUsamp} \Ltwo{ \sampvec_{(i)}}
        + \frac{\delta}{2} \FN{\sampM}^{2} \right) \\
      &- \left( \NN{\varM \sampM}
        + \lambda \Omega_{\numAdd} (\sampM_{\setUsamp}) \right).
    \end{split}
  \end{align}
  Here, both the first and second parenthesized terms are proper lower semi-continuous convex functions, and the objective function is lower bounded.

  \subsection{OPTIMIZATION}
  The proposed problem~\eqref{prob:prop} has a DC structure and can be solved efficiently using the GDPGDC algorithm.
  We map our problem to the general form in~\eqref{prob:DC_general} by setting the components as follows:
  \begin{align}
    f_1(\mat) &= 0, \\
    f_2(\mat) &= \iota_{\setNsampM}(\mat)
          + \lambda \sum_{i \in \setUsamp} \Ltwo{ \vmat{i}}
          + \frac{\delta}{2} \FN{\mat}^{2}, \label{func:f2}\\
    h(\DCw) &= \NN{\DCw_1} + \lambda \Omega_{\numAdd}(\DCw_2), \label{func:h}\\
    \DCw &=
      \begin{bmatrix}
        \DCw_1 \\
        \DCw_2
      \end{bmatrix}
      = \matDC \mat, \
    \matDC =
      \begin{bmatrix}
        \varM \\
        \selUsamp
      \end{bmatrix}, \
    \mat = \sampM,
  \end{align}
  where $\selUsamp \in \realnn{|\setUsamp|}{\numv}$ is a selection matrix that extracts the rows corresponding to the undecided set $\setUsamp$ from a matrix.
  The function $f_2$ groups the avoided vertices constraint, the sparsity-inducing term for undecided vertices, and the regularization term.
  The function $h$ corresponds to the convex terms we aim to maximize: the nuclear norm for the best possible recovery and the sum of the largest row norms to control the number of selected vertices.

  The GDPGDC algorithm, summarized in Algorithm~\ref{algo:GDPG}, involves two main steps that can be solved analytically.
  The first step (line \ref{al_s} in Algorithm~\ref{algo:GDPG}) requires computing $\prox_{\gamma_1^{(t)} f_2}(\cdot)$. Since $f_2$ is separable row-wise, this can be computed for each row $\vmat{i}$ of a given matrix $\mat$ as:
  \begin{align} \label{eq:prox_f2_row}
    &[ \prox_{ \gamma_1^{(t)} f_2 } (\mat) ]_{(i)} = \nonumber \\
    &\begin{cases}
      \zeros, & \text{if } i \in \setNsamp, \\
      \frac{1}{1 + \gamma_1^{(t)} \delta} \max \left(
        0, 1 - \frac{\gamma_1^{(t)} \lambda}{\Ltwo{\vmat{i}}}
      \right) \vmat{i}, & \text{if } i \in \setUsamp, \\
      \frac{1}{1 + \gamma_1^{(t)} \delta} \vmat{i}, & \text{if } i \in \setSamp,
    \end{cases}
  \end{align}
  where $\zeros$ is a zero vector of appropriate dimension.
  For the detail of this calculation, see Appendix~\ref{appendix:calc_prox_f2_row}.

  The second step (line \ref{al_z}) involves $\prox_{\gamma_2^{(t)} h^*}(\cdot)$, which is calculated via Moreau's identity \eqref{eqDCh}.
  The computation of $\prox_{\gamma h}(\DCw)$ can be decomposed into two independent blocks as follows:
  \begin{align}
    \prox_{\gamma h}(\DCw) =
    \begin{bmatrix}
      \prox_{\gamma \NN{\cdot}}(\DCw_1) \\
      \prox_{\gamma \lambda \Omega_{\numAdd}(\cdot)}(\DCw_2)
    \end{bmatrix}.
  \end{align}

  The proximity operator for the first block, $\prox_{\gamma \NN{\cdot}}(\DCw_1)$, is computed as follows:
  \begin{equation}
    \prox_{\gamma \NN{\cdot}}(\DCw_1)
    = \mathbf{U}_{\DCw_1} \mathcal{S}_{\gamma}(\mathbf{\Sigma}_{\DCw_1}) \mathbf{V}_{\DCw_1}^\top, \\
  \end{equation}
  where $\mathbf{U}_{\DCw_1}$ and $\mathbf{V}_{\DCw_1}$ are the left and right singular vectors of $\DCw_1$, respectively, and $\mathcal{S}_{\gamma}(\cdot)$ is the soft-thresholding operator applied to the singular values, defined as: $\mathcal{S}_{\gamma}(\sigma_i) = \max(\sigma_i - \gamma, 0)$, and $\mathcal{S}_{\gamma}(\mathbf{\Sigma}_{\DCw_1})$ is a diagonal matrix with entries $\mathcal{S}_{\gamma}(\sval{\DCw_1}{i})$.

  The second block, $\prox_{\gamma \lambda \Omega_{\numAdd}(\cdot)}(\DCw_2)$, is analytically computable through a two-stage process~\cite{brzyski2019group}.
  First, an optimal threshold is determined from the row $\ell_2$ norms of $\DCw_2$.
  Let $z_i = \Ltwo{[\DCw_2]_{(i)}}$ be the $\ell_2$ norm of the $i$-th row of $\DCw_2$, and let $\mu_{1} \geq \mu_{2} \geq \dots$ be the sorted values of $z_i$ in descending order.
  We find the largest index $k^{*} \in \{ 1, \dots, \numAdd \}$ such that
  \begin{equation}
    \mu_{k^{*}} > \frac{1}{k^{*}} \left( \sum_{j = 1}^{k^{*}} \mu_j - \gamma \lambda \right).
  \end{equation}
  Second, based on the outcome, each row $[\DCw_2]_{(i)}$ is updated.
  If such a $k^{*}$ exists, the update is given by:
  \begin{align}
    &[ \prox_{\gamma \lambda \Omega_{\numAdd}(\cdot)}(\DCw_2) ]_{(i)} \nonumber \\
    &= \max \left( 0, 1 - \frac{ \sum_{j = 1}^{k^{*}} \mu_j - \gamma \lambda }{k^{*} \Ltwo{[\DCw_2]_{(i)}} } \right) [\DCw_2]_{(i)}.
  \end{align}
  If no such $k^{*}$ exists, the output is simply the input matrix, i.e., $\prox_{\gamma \lambda \Omega_{\numAdd}(\cdot)}(\DCw_2) = \DCw_2$.

  \par\kern-4bp
  \begin{rem}[Convergence of the proposed algorithm]
    \rm{
      Since the objective function of Prob.~\eqref{prob:prop} is bounded below and $f_2$ and $h$ are proper lower semi-continuous convex functions, the assumptions in Theorem~\ref{theo:convergence_GDPGDC} are satisfied.
      This ensures that the sequences $\{ \sampM^{(\Step)} \}_{\Step \in \mathbb{N}}$ and $ \{ \DCw^{(\Step)} \}_{\Step \in \mathbb{N}} $ generated by Algorithm~\ref{algo:GDPG} are bounded.
      Therefore, the sequence $\{ \sampM^{(\Step)} \}_{\Step \in \mathbb{N}}$ is guaranteed to converge to a critical point of Prob.~\eqref{prob:prop}.
    }
  \end{rem}

  \par\kern-6bp
  \begin{rem}[Computational complexity]
    \rm{
      The computational cost is determined by each iteration of Algorithm~\ref{algo:GDPG}.
      The computational cost per iteration is $O(NMJ)$.
      For the subspace prior, $J = \sbgenn$; thus, it is $O(\numv \nums \sbgenn)$.
      For the smoothness and stochastic priors, $J = \numv$; thus, it is $O(\numv^2 \nums)$.
    }
  \end{rem}

  \setlength{\textfloatsep}{12pt}

\par\kern-12.3bp
\section{EXPERIMENTS AND RESULTS}
\label{sec:experiments}
    We demonstrated the effectiveness of our method through\linebreak
sampling and recovering experiments across various types\linebreak
of graph signals. All experiments were conducted using\linebreak
MATLAB (R2024b) on a Windows 11 computer with Intel\linebreak
Core i9-14900KF 3.20-GHz processor, 32 GB of RAM, and\linebreak
NVIDIA GeForce RTX 4090. Our method was compared\linebreak
with the following graph signal sampling methods: SP~\cite{anis2016efficient},\linebreak
AVM~\cite{jayawant2022practical}, GSSS~\cite{hara2022gsss}, and ScFGSS~\cite{yamashita2025controlling}. SP and AVM\linebreak
are representative methods for sampling bandlimited graph
\OJDepthColumnBreak{2.089bp}\clearpage
\noindent\begin{minipage}[t][146bp][t]{\columnwidth}\vspace*{-6.6bp}\setlength{\intextsep}{0pt}  \begin{algorithm}[H]
    \caption{GDPGDC-Based Algorithm for Solving~\eqref{prob:prop}}
    \label{algo:GDPG}
    \begin{algorithmic}[1]
    \renewcommand{\algorithmicrequire}{\textbf{Input:}}
    \renewcommand{\algorithmicensure}{\textbf{Output:}}
    \REQUIRE $ \sampM^{(0)}, \DCw^{(0)}, \gamma^{(0)}_1>0, \gamma^{(0)}_2>0$
    \WHILE{a stopping criterion is not met}
    \STATE
      $
        \sampM^{(\Step+1)} \leftarrow
          \prox_{\gamma_1^{(\Step)} f_2} \left(
            \sampM^{(\Step)} + \gamma_{1}^{(\Step)} \matDCT \DCw^{(\Step)}
          \right);
      $
      \label{al_s}
    \STATE
      $
        \DCw^{(\Step+1)} \leftarrow
          \prox_{\gamma_{2}^{(\Step)} h^{*}} \left(
            \DCw^{(\Step)} + \gamma_{2}^{(\Step)} \matDC \sampM^{(\Step + 1)}
          \right);
      $
      \label{al_z}
    \STATE
      $ \Step \leftarrow \Step+1 $;
    \ENDWHILE
    \ENSURE $\sampM^{(\Step)}$
    \end{algorithmic}
  \end{algorithm}
\end{minipage}\par\noindent
signals.
    GSSS is a vertex-wise sampling method for sampling beyond bandlimited graph signals under arbitrary priors.
    ScFGSS is a vertex-wise aggregation sampling method for sampling beyond bandlimited graph signals under arbitrary priors and controlling the number of active vertices without constraints on must-use or avoided vertices, whose convergence is not guaranteed.

    \subsection{SYNTHETIC DATA}
    \subsubsection{SETUP}
    We generated random sensor weighted graphs, which are implemented by k nearest neighbor graphs, whose vertices are randomly distributed in 2-D space $[0, 1] \times [0, 1]$, consisting of $\numv = 256$ vertices by using GSPBox~\cite{perraudin2014gspbox}.
    The size of sampled signal was set as $\nums = 32$.

    We have generated the following types of graph signals for the $i$-th and the largest graph frequency, $\lambda_{i}$ and $\lambda_{\max}$: 
    \vspace{-0.3\baselineskip}
    \begin{itemize}
    \itemsep=0pt \parsep=0pt
    \setlength{\leftskip}{-4mm}
    \setlength{\labelsep}{1mm}
        \item SB prior: Periodic graph spectrum (PGS)~\cite{chen2016representations} signals with the generator response of $\sbgenM$, $\widehat{A}(\lambda_{i}) = \exp(-1.5 \lambda_{i}/\lambda_{\max})$ and $\sbexpd$ with $d_i \sim \mathcal{N}(1,1)$ for any $i \in \{1, \ldots, \sbgenn\}$, $\sbgenn = 16$;
        \item SM prior: Gaussian Markov random field (GMRF)~\cite{gadde2015probabilistic} with the power spectrum $\smps = 0.1 / \left(\lambda_{i} + 0.1\right)$ for any $i = 1, \ldots, \numv$;
        \item ST prior: GMRF with the power spectrum $\stelefsigM = \exp\left(-\left(\left(2 \lambda_{i} - \lambda_{\max}\right)/{\sqrt{\lambda_{\max}}}\right)^2\right)$ for any $i = 1, \ldots, \numv$.
    \end{itemize}

    For all graph types above, we also experimented with noisy sampled signals $\ssign := \gsampsig + \stnoise$, where $\stnoise \in \realM$ is generated as a zero-mean white Gaussian noise with its variance $\sigma^2=0.1$.

    We set the desired upper limit of active vertices $\numz = 32$, and the number of must-use vertices $| \setSamp |$ and the number of avoided vertices $| \setNsamp |$ as $| \setSamp | = | \setNsamp | = 16$.
    We designed the sampling operator $\sampM$ in the following three patterns:\vspace{-4bp}
    \begin{itemize}
    \itemsep=0pt \parsep=0pt
    \setlength{\leftskip}{-4mm}
    \setlength{\labelsep}{1mm}
        \item Design (i): $\setSamp$ is set to $16$ vertices selected by the greedy algorithm proposed in~\cite{hara2022gsss}, and $\setNsamp$ is set to $16$ randomly selected vertices from the remaining vertices not selected in the aforementioned $16$ vertices.
        \item Design (ii): $\setSamp$ is set to $16$ randomly selected vertices from the entire vertices, and $\setNsamp$ is set to $16$ randomly selected vertices from the remaining vertices.
        \item Design (iii): $\setNsamp$ is set to $16$ randomly selected vertices from the first $32$ vertices selected by the greedy algorithm proposed in~\cite{hara2022gsss}, and $\setNsamp$ is set to $16$ randomly
\OJDepthColumnBreak{2.156bp}
\noindent\hspace*{-9.612pt}\begin{minipage}[t][95bp][t]{\columnwidth}\vspace*{-6.49bp}\OJContent{\begin{table}[!t]
    \captionsetup{font=scriptsize} 
    \centering
    \renewcommand{\arraystretch}{1}
    \footnotesize
    \caption{{Parameters for the Proposed Method.}}
    \vspace*{-1mm}
    \label{tab:params}
    {
      \begin{tabular}{c|cccc}
        \toprule
        Prior
        & $\lambda$
        & $\delta$
        & $\gamma_1^{(0)}$
        & $\gamma_2^{(0)}$\\
        \midrule
        Subspace (SB)
        & $1.05$ & $10^{-1}$ & $10^{-3}$ & $10^{-5}$  \\
        Smoothness (SM)
        & $5.1$ & $10^{-1}$ & $10^{-2}$ & $10^{-2}$  \\
        Stochastic (ST)
        & $0.75$ & $10^{-6}$ & $10^{-3}$ & $10^{-5}$ \\
        \bottomrule
      \end{tabular}
    } 
  \end{table}}\end{minipage}\par\noindent
selected vertices from the remaining vertices not selected in the aforementioned $32$ vertices.
    \end{itemize}
    \begingroup\fontsize{10}{11.5}\selectfont

    The parameters for the proposed method are summarized in Table~\ref{tab:params}; $\lambda$ was tuned to target $L$ active vertices, a small $\delta$ was set for the lower boundedness with minimal impact, and small initial stepsizes $\gamma_1^{(0)}, \gamma_2^{(0)}$, which were decreased by $0.01\%$ per iteration, were chosen for stable convergence.
    We defined the initial value $\sampM^{(0)}$ as a matrix with random Gaussian entries and $\DCw^{(0)}$ as $\DCw^{(0)} = [\varM^{\top} \ \selUsamp^{\top}]^{\top} \sampM^{(0)}$.
    The stopping criterion was set as $\|\sampM^{(\Step + 1)} - \sampM^{(\Step)}\|_{F}/\|\sampM^{(\Step)}\|_{F} \leq 10^{-5}$.
    The parameter settings for the existing methods followed the descriptions in the respective papers, and a graph filter used in GSSS was set as an identity matrix.

    For sampling graph signals under the smoothness prior, we used the spectral response $\smopefil$ of the smoothness operator $\smopeM$ following as
        $F(\lambda_i) = {\lambda_i}/{\lambda_\mathrm{max}} + 0.1$.

    For the quantitative evaluations, we used MSE: $\mathrm{MSE} = \|\tilde{\gsig} - \gsig\|^{2}_2/\numv$, where $\tilde{\gsig}$ and $\gsig$ are the recovered and original graph signals, respectively.
    As a note, smaller values of the MSE indicate better recovery accuracy.

    \par\kern11bp
    \subsubsection{RESULTS AND DISCUSSION}

    Table~\ref{table:mse} represents the averaged MSEs in decibels between original graph signals and recovered graph signals generated by each method obtained from 20 independent runs for each graph signal type.
    Table~\ref{table:numS} represents the average number of active vertices by the sampling operators designed by the proposed method for 20 independent runs.
    Fig.~\ref{fig:result} visualizes examples of the original and recovered graph signals.
    Fig.~\ref{fig:sampM} visualizes the sampling operators designed by the proposed method for the PGS graph signal illustrated in Fig.~\ref{fig:result}.

    The results shown in Table~\ref{table:mse}, comparing the results of SP and AVM, specialized for bandlimited graph signals, with those of GSSS, ScFGSS, and the proposed method, applicable to beyond bandlimited graph signals, show that the latter methods are superior in all cases.
    This indicates the effectiveness to be applicable to beyond bandlimited graph signals based on the generalized sampling theory.
    
    Furthermore, comparing the results of GSSS, which performs a vertex-wise sampling, with those of ScFGSS and the proposed method, which perform vertex-wise aggregation sampling without and with the constraints on prior knowledge of vertices, respectively, in most cases, especially in cases with noise, ScFGSS and the proposed method achieved better results.
    This indicates the effectiveness of aggregation sampling approaches.
    In the noiseless case of the subspace prior, GSSS achieved the best result; however, GSSS, ScFGSS, \par\endgroup\OJDepthColumnBreak{1.439bp}\twocolumn[{\begin{minipage}[t][326bp][t]{\textwidth}\setlength{\columnwidth}{\textwidth}\OJContent{\begin{table*}[!t]
    \captionsetup{font=scriptsize} 
    \centering
    \footnotesize
    \caption{{Average MSEs in Decibel of the Recoveries for 20 Independent Runs on each Method.}}
    \vspace{-1mm}
    \label{table:mse}
    \renewcommand{\arraystretch}{0.95}
    {
      \tabcolsep = 7pt 
      \begin{tabular}{c|ccccccc}
        \toprule
        Signal Prior
        & SP~\cite{anis2016efficient} 
        & AVM~\cite{jayawant2022practical} 
        & GSSS~\cite{hara2022gsss} 
        & ScFGSS~\cite{yamashita2025controlling}
        & \textbf{Design (i)}
        & \textbf{Design (ii)}
        & \textbf{Design (iii)} \\
        \midrule
        \vspace{0mm}
        SB (PGS)
        & $-19.672$ & $-18.474$ & $\mathbf{-612.924}$ & $-604.539$ & $-606.181$ & $\underline{-608.050}$ & $-601.380$\\
        + noise
        & $-14.896$ & $-13.730$ & $-27.669$ & $\mathbf{-38.753}$ & $\underline{-37.838}$ & $-22.354$ & $-16.717$ \\
        SM (GMRF)
        & $-23.630$ & $-6.125$ & $-27.559$ & $\mathbf{-27.739}$ & $\underline{-27.679}$ & $-27.395$ & $-27.392$\\ \vspace{0mm}
        + noise
        & $-18.003$ & $-4.600$ & $-22.608$ & $\underline{-26.960}$ & $\mathbf{-27.374}$ & $-26.428$ & $-25.900$\\ \vspace{0mm}
        ST (GMRF)
        & $-20.758$ & $-9.036$ & $-25.243$ & $-26.235$ & $\mathbf{-26.705}$ & $\underline{-26.617}$ & $-26.572$\\ \vspace{0mm}
        + noise
        & $-15.137$ & $-6.158$ & $-23.972$ & $\mathbf{-25.933}$ & $\underline{-25.553}$ & $-25.373$ & $-25.368$\\
        \bottomrule
      \end{tabular}
      \vspace{-3mm}
    } 
  \end{table*}\vspace{36bp}
\def\resultA{$-21.992$}
\def\resultB{$-23.603$}
\def\resultC{$\mathbf{-622.212}$}
\def\resultE{$\underline{-621.297}$}
\def\resultF{$-614.793$}
\def\resultG{$-619.431$}
\def\resultH{$-601.447$}

\def\resultNoiseA{$-8.676$}
\def\resultNoiseB{$-22.913$}
\def\resultNoiseC{$-26.505$}
\def\resultNoiseD{$-25.148$}
\def\resultNoiseE{$\underline{-43.229}$}
\def\resultNoiseF{$\mathbf{-49.059}$}
\def\resultNoiseG{$-34.314$}
\def\resultNoiseH{$-30.174$}

\begin{figure*}[t]
    \centering
    \captionsetup{font=scriptsize} 
    \begin{subfigure}[t]{20.2mm}
        \includegraphics[width=20.2mm]{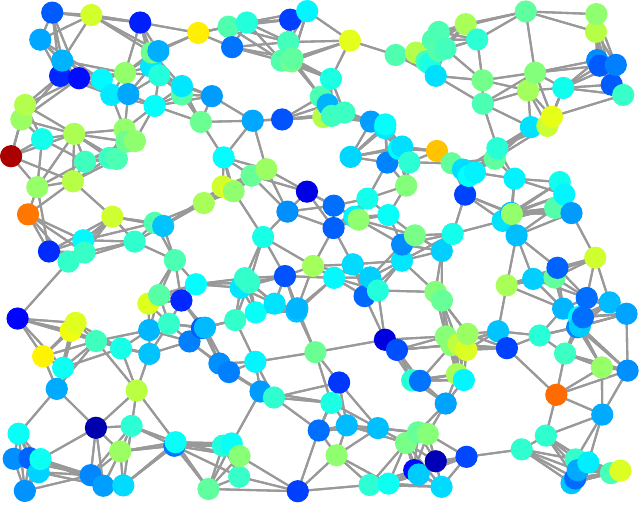}
        \vspace*{-5.5mm}
        \subcaption{Original\\[0mm]MSE (dB)}
        \vspace*{-2mm}
        \label{fig:ex_original}
    \end{subfigure}
    \hfill
    \begin{subfigure}[t]{20.2mm}
        \includegraphics[width=20.2mm]{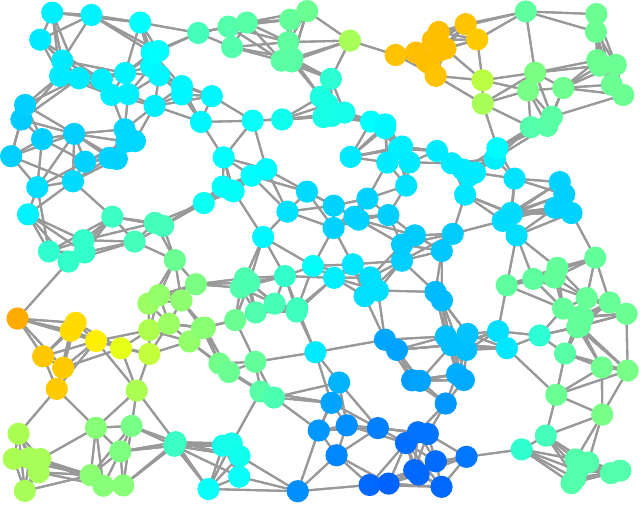}
        \vspace*{-5.5mm}
        \subcaption{SP~\cite{anis2016efficient}\\[0mm] \resultA}
        \vspace*{-2mm}
        \label{fig:ex_sp}
    \end{subfigure}
    \hfill
    \begin{subfigure}[t]{20.2mm}
        \includegraphics[width=20.2mm]{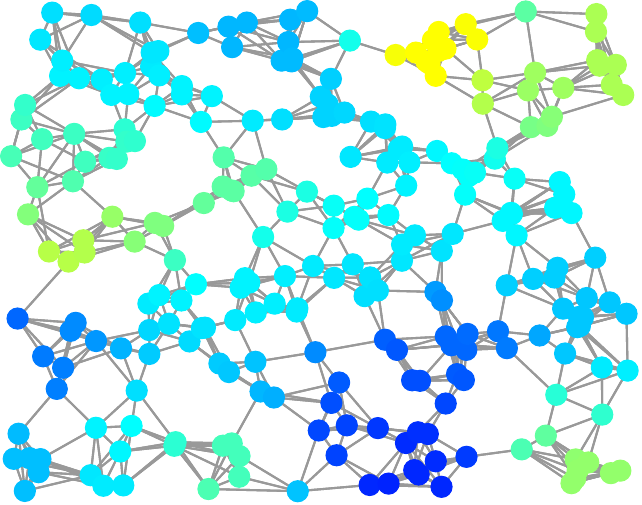}
        \vspace*{-5.5mm}
        \subcaption{AVM~\cite{jayawant2022practical}\\[0mm] \resultB}
        \vspace*{-2mm}
        \label{fig:ex_avm}
    \end{subfigure}
    \hfill
    \begin{subfigure}[t]{20.2mm}
        \includegraphics[width=20.2mm]{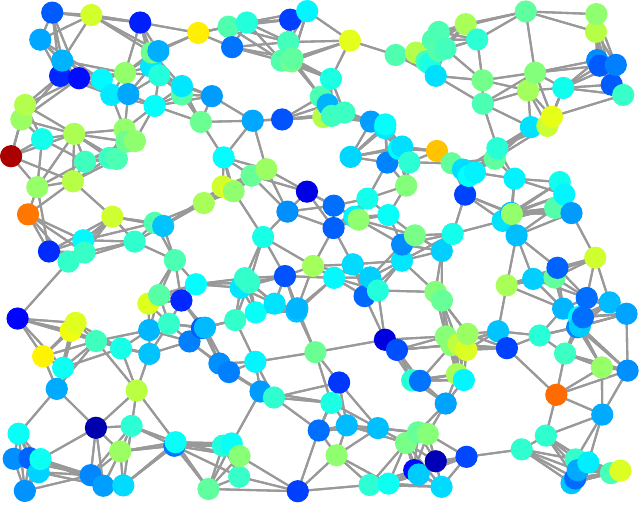}
        \vspace*{-5.5mm}
        \subcaption{GSSS~\cite{hara2022gsss}\\[0mm] \resultC}
        \vspace*{-2mm}
        \label{fig:ex_gsss}
    \end{subfigure}
    \hfill
    \begin{subfigure}[t]{20.2mm}
        \includegraphics[width=20.2mm]{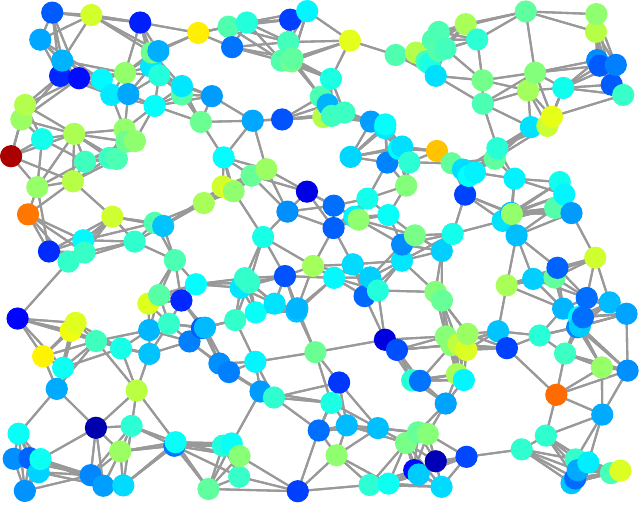}
        \vspace*{-5.5mm}
        \subcaption{ScFGSS~\cite{yamashita2025controlling}\\[0mm] \resultE}
        \vspace*{-2mm}
        \label{fig:ex_scfgss}
    \end{subfigure}
    \hfill
    \begin{subfigure}[t]{20.2mm}
        \includegraphics[width=20.2mm]{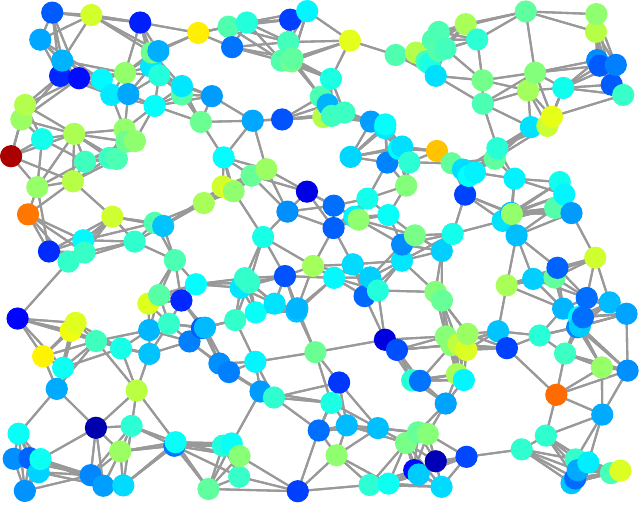}
        \vspace*{-5.5mm}
        \subcaption{\textbf{Design (i)}\\[0mm] \resultF}
        \vspace*{-2mm}
        \label{fig:ex_proposed1}
    \end{subfigure}
    \hfill
    \begin{subfigure}[t]{20.2mm}
        \includegraphics[width=20.2mm]{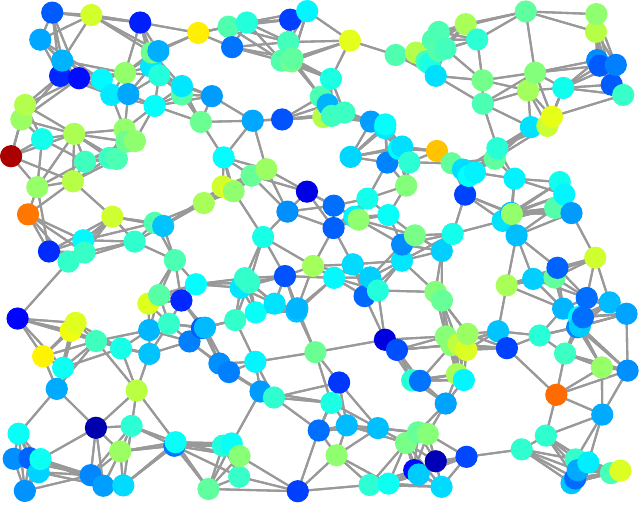}
        \vspace*{-5.5mm}
        \subcaption{\textbf{Design (ii)}\\[0mm] \resultG}
        \vspace*{-2mm}
        \label{fig:ex_proposed2}
    \end{subfigure}
    \hfill
    \begin{subfigure}[t]{20.2mm}
        \includegraphics[width=20.2mm]{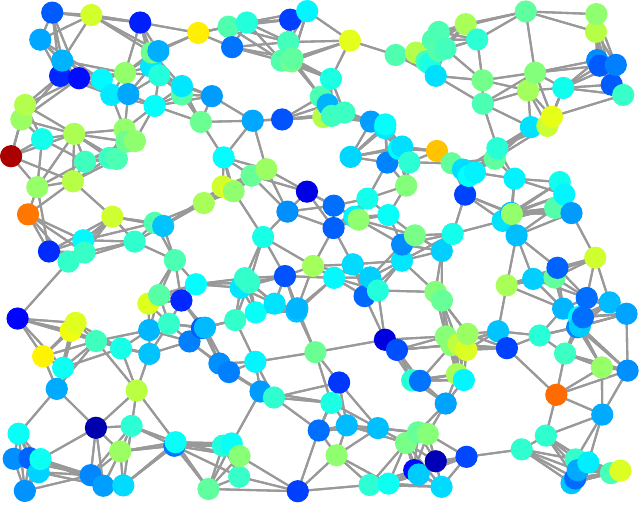}
        \vspace*{-5.5mm}
        \subcaption{\textbf{Design (iii)}\\[0mm] \resultH}
        \vspace*{-2mm}
        \label{fig:ex_proposed3}
    \end{subfigure}
    \begin{subfigure}[t]{4mm}
        \includegraphics[width=4mm]{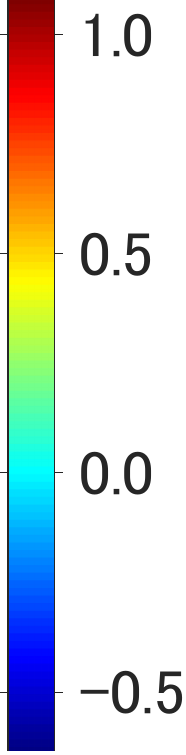}
        \vspace*{-5.5mm}
        \label{fig:colorbar}
        \vspace*{-2mm}
    \end{subfigure}

    \begin{subfigure}[t]{20.2mm}
        \hspace{20.2mm}
    \end{subfigure}
    \hfill
    \begin{subfigure}[t]{20.2mm}
        \includegraphics[width=20.2mm]{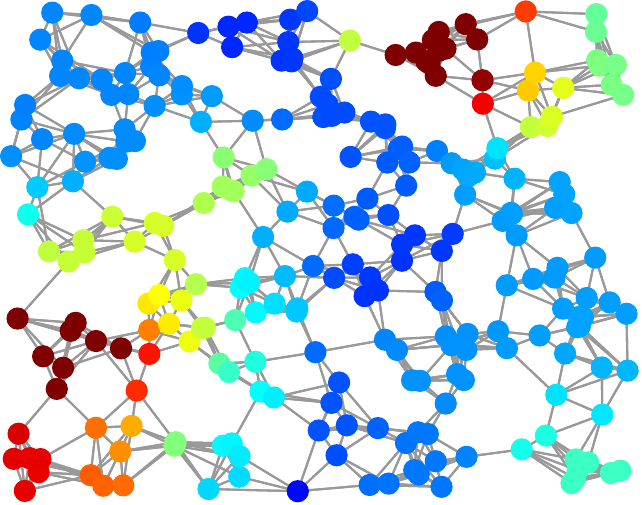}
        \vspace*{-5.5mm}
        \subcaption{SP~\cite{anis2016efficient}\\[0mm] \resultNoiseA}
        \label{fig:ex_noise_sp}
    \end{subfigure}
    \hfill
    \begin{subfigure}[t]{20.2mm}
        \includegraphics[width=20.2mm]{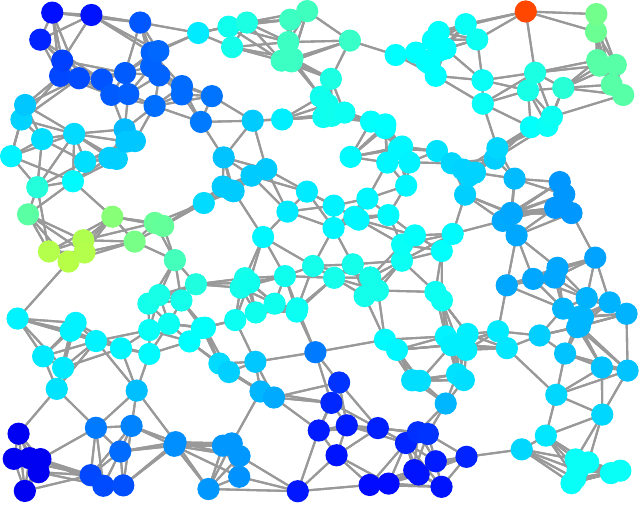}
        \vspace*{-5.5mm}
        \subcaption{AVM~\cite{jayawant2022practical}\\[0mm] \resultNoiseB}
        \label{fig:ex_noise_avm}
    \end{subfigure}
    \hfill
    \begin{subfigure}[t]{20.2mm}
        \includegraphics[width=20.2mm]{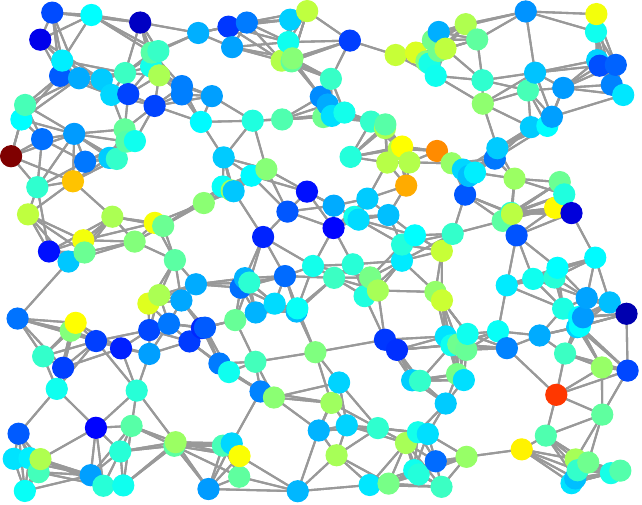}
        \vspace*{-5.5mm}
        \subcaption{GSSS~\cite{hara2022gsss}\\[0mm] \resultNoiseC}
        \label{fig:ex_noise_gsss}
    \end{subfigure}
    \hfill
    \begin{subfigure}[t]{20.2mm}
        \includegraphics[width=20.2mm]{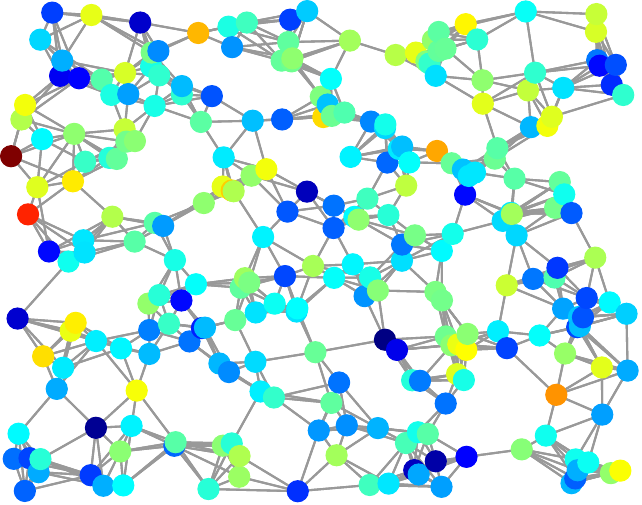}
        \vspace*{-5.5mm}
        \subcaption{ScFGSS~\cite{yamashita2025controlling}\\[0mm] \resultNoiseE}
        \label{fig:ex_noise_scfgss}
    \end{subfigure}
    \hfill
    \begin{subfigure}[t]{20.2mm}
        \includegraphics[width=20.2mm]{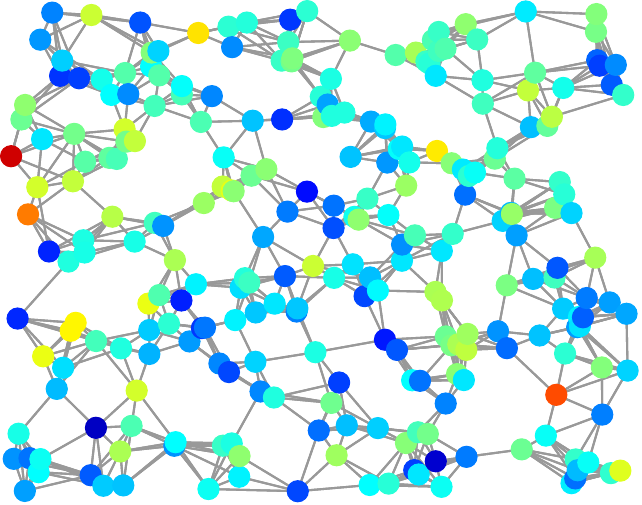}
        \vspace*{-5.5mm}
        \subcaption{\textbf{Design (i)}\\[0mm] \resultNoiseF}
        \label{fig:ex_noise_proposed1}
    \end{subfigure}
    \hfill
    \begin{subfigure}[t]{20.2mm}
        \includegraphics[width=20.2mm]{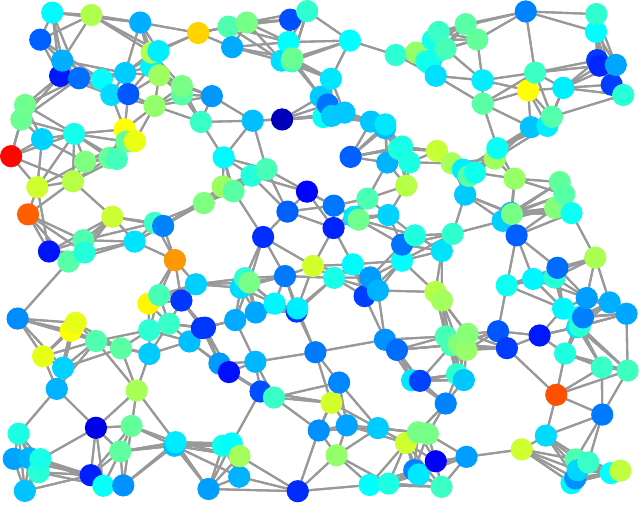}
        \vspace*{-5.5mm}
        \subcaption{\textbf{Design (ii)}\\[0mm] \resultNoiseG}
        \label{fig:ex_noise_proposed2}
    \end{subfigure}
    \hfill
    \begin{subfigure}[t]{20.2mm}
        \includegraphics[width=20.2mm]{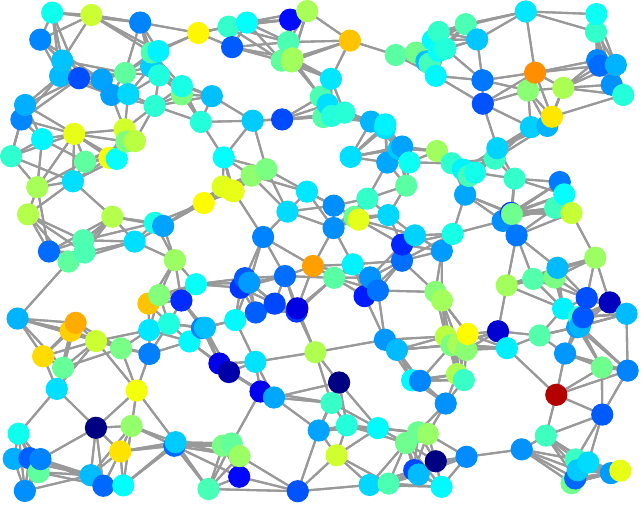}
        \vspace*{-5.5mm}
        \subcaption{\textbf{Design (iii)}\\[0mm] \resultNoiseH}
        \label{fig:ex_noise_proposed3}
    \end{subfigure}
    \begin{subfigure}[t]{4mm}
        \includegraphics[width=4mm]{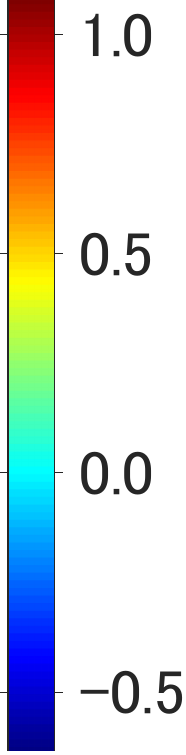}
        \vspace*{-5.5mm}
        \label{fig:colorbar_noise}
    \end{subfigure}

    \vspace{-3mm}
    \captionsetup{font=scriptsize}
    \caption{{
        An example of sampling and recovery experiment on a PGS graph signal on a random sensor graph. 
        Fig.~(\subref*{fig:ex_original}) shows the original PGS signal. 
        Figs.~(\subref*{fig:ex_sp})-(\subref*{fig:ex_proposed3}) show the recovered signals from the noiseless sampled signals,
        and Figs.~(\subref*{fig:ex_noise_sp})-(\subref*{fig:ex_noise_proposed3}) show the recovered signals from the noisy sampled signals.
        The results in MSE are expressed in decibels, where smaller values indicate better recovery accuracy.
        The best and second best results in each case are in bold and underlined, respectively.
        }}
    \vspace{-5mm}
    \label{fig:result}
\end{figure*}}\end{minipage}}]
\noindent\begin{minipage}[t][320bp][t]{\columnwidth}\vspace*{-6.6bp}\OJContent{%
\begin{minipage}[t][97.46bp][t]{\columnwidth}\begin{table}[!t]
    \captionsetup{font=scriptsize} 
    \centering
    \footnotesize
    \caption{{Average Number of Active Vertices by the Proposed Method for 20 Independent Runs.}}
    \vspace{-1mm}
    \label{table:numS}
    \renewcommand{\arraystretch}{0.9}
    {
      \tabcolsep = 5pt 
      \begin{tabular}{c|ccc}
        \toprule
        Signal Prior
        & Design (i)
        & Design (ii)
        & Design (iii) \\
        \midrule
        \vspace{0mm}
        SB (PGS)
        & $24.20$ & $33.60$ & $31.50$ \\ \vspace{0mm}
        SM (GMRF)
        & $30.75$ & $31.10$ & $31.00$ \\ \vspace{0mm}
        ST (GMRF)
        & $31.85$ & $31.85$ & $32.00$ \\
        \bottomrule
      \end{tabular}
      \vspace{-4mm}
    } 
  \end{table}\end{minipage}\par
\begin{table}[!t]
    \captionsetup{font=scriptsize} 
    \centering
    \renewcommand{\arraystretch}{1}
    \footnotesize
    \caption{{Parameters for the Proposed Method.}}
    \vspace*{-1mm}
    \label{tab:params_real}
    {
      \begin{tabular}{c|cccc}
        \toprule
        Prior
        & $\lambda$
        & $\delta$
        & $\gamma_1^{(0)}$
        & $\gamma_2^{(0)}$\\
        \midrule
        Smoothness (SM)
        & $1.2$ & $10^{-6}$ & $10^{-2}$ & $10^{-5}$  \\
        Stochastic (ST)
        & $0.33$ & $10^{-6}$ & $10^{-2}$ & $10^{-5}$ \\
        \bottomrule
      \end{tabular}
    } 
    \vspace*{-2mm}
  \end{table}\vspace{46.971bp}\begin{figure}[t]
    \centering
    \captionsetup{font=scriptsize} 
    \begin{subfigure}[t]{24.2mm}
        \includegraphics[width=24.2mm]{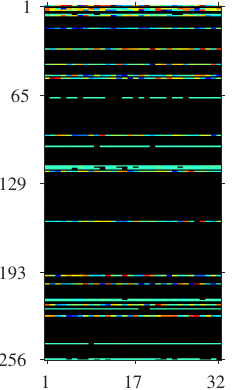}
        \vspace*{-6.5mm}
        \subcaption{\textbf{Design (i)}}
        \label{fig:sampM_proposed1}
    \end{subfigure}
    \hfill
    \begin{subfigure}[t]{24.2mm}
        \includegraphics[width=24.2mm]{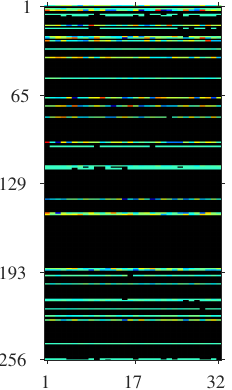}
        \vspace*{-6.5mm}
        \subcaption{\textbf{Design (ii)}}
        \label{fig:sampM_proposed2}
    \end{subfigure}
    \hfill
    \begin{subfigure}[t]{24.2mm}
        \includegraphics[width=24.2mm]{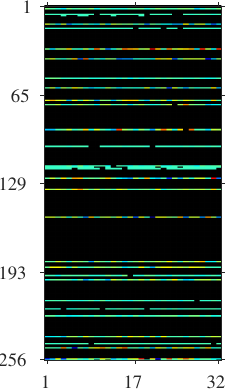}
        \vspace*{-6.5mm}
        \subcaption{\textbf{Design (iii)}}
        \label{fig:sampM_proposed3}
    \end{subfigure}
    \hspace{0.2mm}
    \raisebox{3mm}{\begin{subfigure}[t]{5.8mm}
        \includegraphics[width=5.8mm]{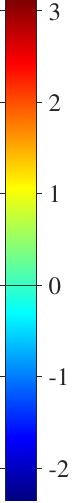}
        \label{fig:colorbar_sampM}
    \end{subfigure}}
    
    \vspace{-3mm}
    \caption{{
        Sampling operators designed by the proposed method for the graph signal in Fig.~\ref{fig:result}.
        Zero values are shown in black.
        }} 
    \label{fig:sampM}
\end{figure}}\end{minipage}\OJColumnBreak\begingroup\setlength{\parskip}{0pt plus1pt}\noindent
and the proposed method all achieved very good results with MSEs exceeding $-600$ dB ($10^{-31}$).

    The results of ScFGSS and Design (i) are equivalent in many cases.
    In the proposed method, there are constraints on must-use and avoided vertices, while in ScFGSS, there are no such constraints, allowing for a more flexible design of the sampling operator.
    In addition, in the proposed method, the number of active vertices is controlled by a DC penalty term in the objective function, making it difficult to strictly ensure that the number is equal to $32$ as shown in Table~\ref{table:numS}; however, in ScFGSS, this number is strictly controlled to be exactly $32$.
    The fact that the proposed method can achieve results equivalent to those of ScFGSS under these conditions demonstrates the effectiveness of the proposed method that can handle constraints on must-use and avoided vertices.

    Comparing the results of Designs (i)-(iii), the results of Designs (ii) and (iii) were less accurate than that of Design (i) due to their randomness or lack of important vertices in their setting of must-use and avoided vertices.
    However, they still outperformed or were comparable to the existing methods in most cases.
    This suggests that the proposed method is relatively robust to the settings of the must-use and avoided vertices.

    In Fig.~\ref{fig:result}, an example of the recovery results is visualized, which also shows that high accurate sampling and recovery is achieved by the proposed method.
    In Fig.~\ref{fig:sampM}, an example of the sampling operators designed by the proposed method is shown, which also shows that the sampling operators designed \par\endgroup\OJDepthColumnBreak{2.156bp}\clearpage
\noindent\begin{minipage}[t][133bp][t]{\columnwidth}\vspace*{2.32bp}\OJContent{\begin{figure}[t]
    \centering
    \captionsetup{font=scriptsize} 
    \begin{subfigure}[t]{28.0mm}
        \includegraphics[width=28.0mm]{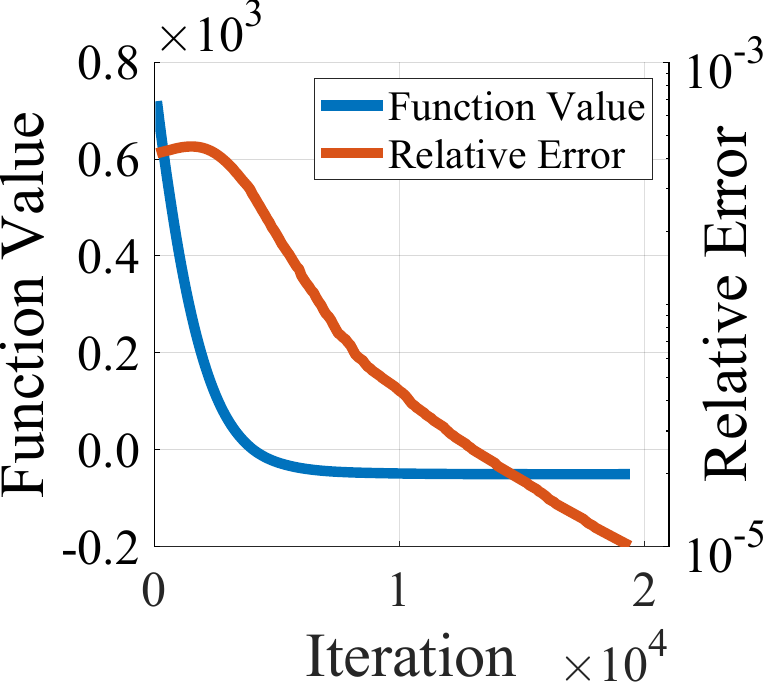}
        \vspace*{-5.5mm}
        \subcaption{Subspace Prior}
        \vspace*{-2mm}
        \label{fig:conv_subspace}
    \end{subfigure}
    \hfill
    \begin{subfigure}[t]{28.0mm}
        \includegraphics[width=28.0mm]{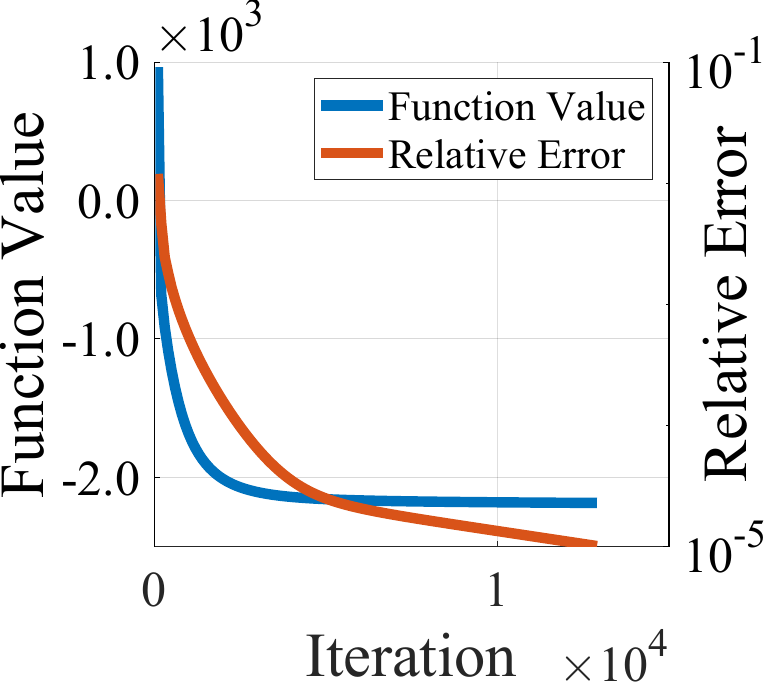}
        \vspace*{-5.5mm}
        \subcaption{Smoothness Prior}
        \vspace*{-2mm}
        \label{fig:conv_smoothness}
    \end{subfigure}
    \hfill
    \begin{subfigure}[t]{28.0mm}
        \includegraphics[width=28.0mm]{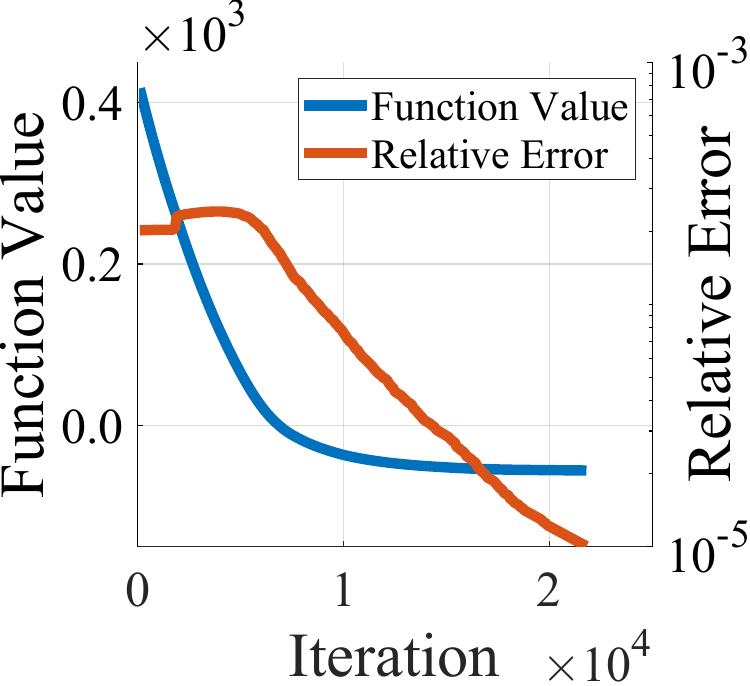}
        \vspace*{-5.5mm}
        \subcaption{Stochastic Prior}
        \vspace*{-2mm}
        \label{fig:conv_stochastic}
    \end{subfigure}
    \vspace{-3mm}
    \captionsetup{font=scriptsize}
    \caption{{
        Examples of convergence behavior of the proposed method (Design (i)): objective function values and relative errors.}
        }
    \label{fig:convergence}
\end{figure}}\end{minipage}\par\noindent
by the proposed method can control the number of active vertices.
    Fig.~\ref{fig:convergence} shows examples of the convergence behavior of the proposed method (Design (i)) for each prior that confirm the convergences of our method.

    \subsection{REAL-WORLD DATA}
    \subsubsection{SETUP}
    We used a dataset of monthly average temperatures from 2015 to 2024 at $\numv = 110$ stations in Switzerland~\cite{MeteoSwissData}.
    The data were originally in Celsius, so we normalized them so that the maximum absolute value of the average temperature for each month would be~$1$. 
    The data was divided into two periods.
    The data from 2015 to 2019 was used to construct the graph via a graph learning method~\cite{dong2016learning} for the smoothness prior and to estimate the signal covariance matrix~\cite{perraudin2017stationary} for the stochastic prior. 
    The remaining data served as the graph signals for the sampling and recovery experiments. 

    The size of sampled signal was set to $\nums = 28$. 
    We also conducted experiments where zero-mean white Gaussian noise with a variance of $\sigma^2=0.05$ was added to the sampled signals.
    For experiments assuming a smoothness prior, the smoothness operator $\smopeM$ was defined by its spectral response $F(\lambda_i) = {\lambda_i}/{\lambda_\mathrm{max}} + 0.1$.

    We set the desired upper limit of active vertices to $\numz = 28$.
    The number of must-use and avoided vertices were defined as $| \setSamp | = | \setNsamp | = 14$.
    We employed the same vertex selection strategy as Design~(i) described in the synthetic data experiments.
    The parameters for our method, which were selected with the same approach as in the synthetic data experiment, are summarized in Table~\ref{tab:params_real}, and the stepsizes $\gamma_1^{(t)}$ and $\gamma_2^{(t)}$ were decreased by $0.01\%$ at each iteration.
    The initial values, $\sampM^{(0)}$ and $\DCw^{(0)}$, and the stopping criterion were set in the same manner as in the synthetic data experiments.

    The performance of our method was compared against the same existing methods from the synthetic data experiments, using the average MSE as the evaluation metric.

    \subsubsection{RESULTS}

    Table~\ref{table:mse_real} shows the average MSEs in decibels between original and recovered graph signals.
    The methods proposed in SP and AVM do not consider the difference of the priors, thus their results are only listed in the smoothness prior rows.
    Fig.~\ref{fig:result_real} visualizes examples of the original and recovered graph signals of the average temperature across Switzerland \OJDepthColumnBreak{1.443bp}
\noindent\begin{minipage}[t][275bp][t]{\columnwidth}\vspace*{-6.49bp}\OJContent{\begin{table}[!t]
    \captionsetup{font=scriptsize} 
    \centering
    \footnotesize
    \caption{{
      Average MSEs in Decibel of the Recoveries for Monthly Average Temperature at $110$ Stations in Switzerland.
    }}
    \vspace{-1mm }
    \label{table:mse_real}
    \renewcommand{\arraystretch}{1.0}
    {
      \tabcolsep = 4.7pt
      \begin{tabular}{c|ccccc}
        \toprule
        Prior
        & SP
        & AVM
        & GSSS
        & ScFGSS
        & \textbf{Ours} \\
        \midrule
        \vspace{0mm}
        SM
        & $-29.429$ & $-29.111$ & $-42.689$ & $\mathbf{-44.765}$ & $-44.065$\\ \vspace{0mm}
        + noise
        & $-18.833$ & $8.075$ & $-33.751$ & $-40.807$ & $\mathbf{-40.842}$\\ \vspace{0mm}
        ST
        & $-$ & $-$ & $-48.247$ & $-67.708$ & $\mathbf{-68.279}$\\ \vspace{0mm}
        + noise
        & $-$ & $-$ & $-39.838$ & $-61.148$ & $\mathbf{-64.416}$\\
        \bottomrule
      \end{tabular}
      \vspace{-3mm}
    } 
  \end{table}

\vspace{11.8bp}
\def\resultAr{$-22.169$}
\def\resultBr{$-21.982$}
\def\resultCr{$-47.159$}
\def\resultDr{$-64.423$}
\def\resultEr{$\mathbf{-67.684}$}

\begin{figure}[t!]
    \centering
    \captionsetup{font=scriptsize} 
    \begin{subfigure}[t]{26.3mm}
        \includegraphics[width=26.3mm]{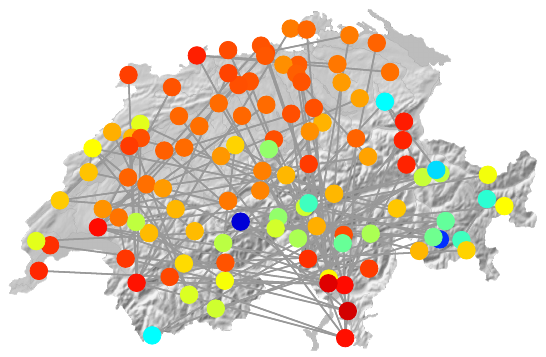}
        \vspace*{-5.5mm}
        \subcaption{Original\\[0mm]MSE (dB)}
        \vspace*{-2mm}
        \label{fig:ex_original_real}
    \end{subfigure}
    \hfill
    \begin{subfigure}[t]{26.3mm}
        \includegraphics[width=26.3mm]{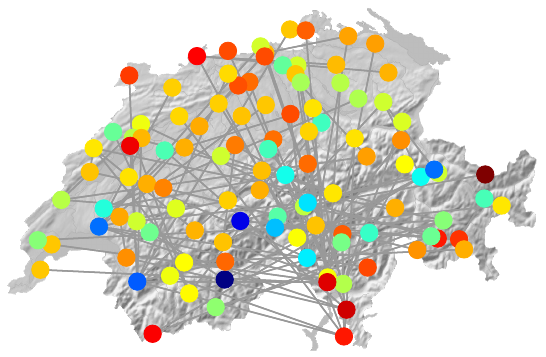}
        \vspace*{-5.5mm}
        \subcaption{SP~[20]\\[0mm] \resultAr}
        \vspace*{-2mm}
        \label{fig:ex_sp_real}
    \end{subfigure}
    \hfill
    \begin{subfigure}[t]{26.3mm}
        \includegraphics[width=26.3mm]{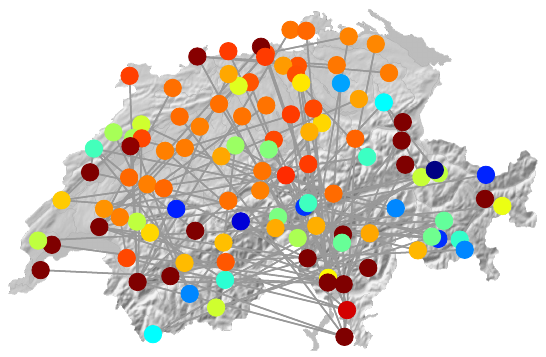}
        \vspace*{-5.5mm}
        \subcaption{AVM~[26]\\[0mm] \resultBr}
        \vspace*{-2mm}
        \label{fig:ex_avm_real}
    \end{subfigure}
    \begin{subfigure}[t]{4.0mm}
        \includegraphics[width=4.0mm]{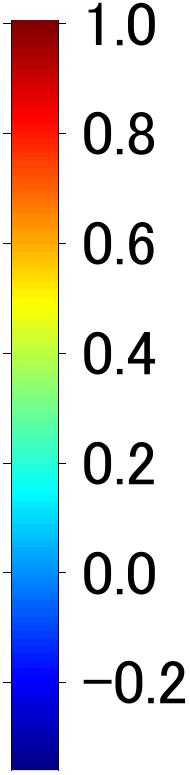}
        \vspace*{-5.5mm}
        \label{fig:colorbar_real}
    \end{subfigure}

    \vspace{0mm}
    \begin{subfigure}[t]{26.3mm}
        \includegraphics[width=26.3mm]{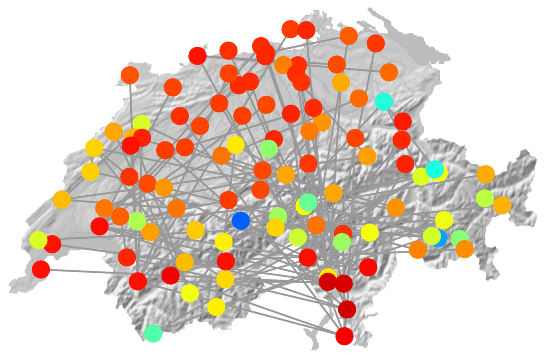}
        \vspace*{-5.5mm}
        \subcaption{GSSS~[33]\\[0mm] \resultCr}
        \label{fig:ex_gsss_real}
    \end{subfigure}
    \hfill
    \begin{subfigure}[t]{26.3mm}
        \includegraphics[width=26.3mm]{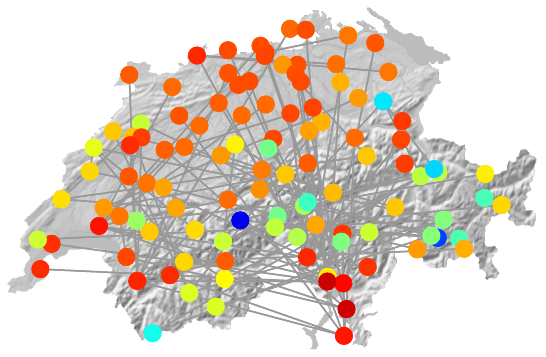}
        \vspace*{-5.5mm}
        \subcaption{ScFGSS~[37]\\[0mm] \resultDr}
        \label{fig:ex_scfgss_real}
    \end{subfigure}
    \hfill
    \begin{subfigure}[t]{26.3mm}
        \includegraphics[width=26.3mm]{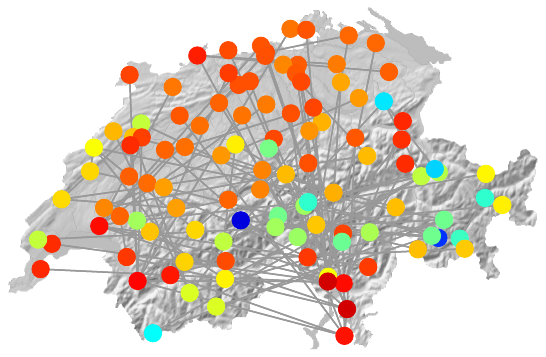}
        \vspace*{-5.5mm}
        \subcaption{\textbf{Ours}\\[0mm] \resultEr}
        \label{fig:ex_proposed1_real}
    \end{subfigure}
    \begin{subfigure}[t]{4.0mm}
        \includegraphics[width=4.0mm]{04withoutNoise/colorbar.pdf}
        \vspace*{-5.5mm}
        \label{fig:colorbar_real2}
    \end{subfigure}
    
    \vspace{-3mm}
    \captionsetup{font=scriptsize}
    \caption{{
        Signal recovery results for the average temperature of Switzerland in October with assuming the stochastic prior.
        Fig.~(\subref*{fig:ex_original_real}) shows the original temperature.
        Figs.~(\subref*{fig:ex_sp_real})-(\subref*{fig:ex_proposed1_real}) show the recovered temperatures.
        The best result is in bold.
        }}
    \vspace{-1mm}
    \label{fig:result_real}
\end{figure}}\end{minipage}\par\noindent
in October from 2020 to 2024 with assuming the stochastic prior.
    Both results demonstrate that the proposed method achieves superior or comparable performance to the existing methods in all cases.
    This indicates the effectiveness of the proposed method even in real-world data experiments.
\par\kern9bp
\section{CONCLUSION}
\label{sec:conclusion}
We proposed a method to design a sampling operator to achieve vertex-wise aggregation sampling for beyond bandlimited graph signals.
The design was formulated as an optimization problem with a rank constraint to achieve the best possible recovery based on the generalized sampling theory, a constraint to control the number of active vertices, and a constraint to handle prior knowledge of sets of must-use and avoided vertices.
The problem was transformed into a DC optimization problem by relaxing the rank and controlling the number of active vertices constraints using a nuclear norm and a DC penalty term, respectively.
To solve it, we developed a solver based on the GDPGDC algorithm, which guarantees convergence to a critical point.
The experiments demonstrated the effectiveness of our method.
\allowdisplaybreaks[4]
\par\kern-9bp
\section*{APPENDIX}
\subsection{THE CALCULATION OF \texorpdfstring{\eqref{eq:prox_f2_row}}{(24)}}
\label{appendix:calc_prox_f2_row}
    Since $f_2$ described in \eqref{func:f2} is separable row-wise, each row of $\prox_{\gamma f_2} (\mat)$ can be represented as follows:
    \begin{align}
        \label{eq:appendix_prox_f2_row_1}
        &\left[\prox_{\gamma f_2} (\mat)\right]_{(i)} \nonumber \\
            &=
                \argmin_{\vmatt{i}} \left\{
                \alpha \iota_{\zeros}(\vmatt{i})
                + \beta \lambda \Ltwo{\vmatt{i}} 
                + \frac{\delta}{2} \FN{\vmatt{i}}^{2} \right. \nonumber \displaybreak[4] \\ 
            &\quad\quad\quad\quad\quad\quad\quad\quad\quad\quad\quad\left.
                + \frac{1}{2\gamma} \FN{\vmat{i} - \vmatt{i}}^{2}
            \right\},
    \end{align}
    \abovedisplayskip=7.5pt \belowdisplayskip=7.5pt
    \abovedisplayshortskip=3pt \belowdisplayshortskip=7.5pt
    where $\alpha = 1$ if $i \in \setNsamp$ and $\alpha = 0$ otherwise, and $\beta = 1$ if $i \in \setUsamp$ and $\beta = 0$ otherwise.
    We consider the following cases depending on which set the vertex $i$ belongs to.
    \begin{itemize}
    \itemsep=0pt \parsep=0pt
    \setlength{\leftskip}{-4mm}
    \setlength{\labelsep}{1mm}
        \item $i \in \setNsamp$: The indicator function $\iota_{\zeros}(\vmatt{i})$ enforces
            \begin{align}
                \label{eq:appendix_prox_f2_row_setN_solution}
                \left[\prox_{\gamma f_2} (\mat)\right]_{(i)} 
                    = \zeros.
            \end{align}
        \item $i \in \setUsamp$: \eqref{eq:appendix_prox_f2_row_1} can be written as
            \begin{align}
                \label{eq:appendix_prox_f2_row_setU}
                &\hspace{-4mm}
                    \left[\prox_{\gamma f_2} (\mat)\right]_{(i)} \nonumber \\
                &\hspace{-4mm} = \resizebox{0.88\columnwidth}{!}{$\displaystyle
                    \argmin_{\vmatt{i}} \left\{
                        \lambda \Ltwo{\vmatt{i}} 
                        + \frac{\delta}{2} \FN{\vmatt{i}}^{2}
                        + \frac{1}{2\gamma} \FN{\vmat{i} - \vmatt{i}}^{2}
                    \right\} $}.
            \end{align}
        This is the proximity operator of the sum of the $\ell_2$ norm and the squared $\ell_2$ norm, and its solution is given by
            \begin{align}
                \label{eq:appendix_prox_f2_row_setU_solution}
                \hspace*{-4mm}\left[\prox_{\gamma f_2} (\mat)\right]_{(i)}
                    = \frac{1}{1 + \gamma \delta} 
                        \max \left\{0, 1 - \frac{\gamma \lambda}{\Ltwo{\vmat{i}}}\right\} \vmat{i}.
            \end{align}
        \item $i \in \setSamp$: \eqref{eq:appendix_prox_f2_row_1} can be written as
            \begin{align}
                \label{eq:appendix_prox_f2_row_setC}
                &\left[\prox_{\gamma f_2} (\mat)\right]_{(i)} \nonumber \\
                    &= \argmin_{\vmatt{i}} \left\{
                        \frac{\delta}{2} \FN{\vmatt{i}}^{2}
                        + \frac{1}{2\gamma} \FN{\vmat{i} - \vmatt{i}}^{2}
                    \right\}.
            \end{align}
        This is the proximity operator of the squared $\ell_2$ norm, and its solution is given by
            \begin{align}
                \label{eq:appendix_prox_f2_row_setC_solution}
                \left[\prox_{\gamma f_2} (\mat)\right]_{(i)} 
                    &= \frac{1}{1 + \gamma \delta} \vmat{i}.
            \end{align}
    \end{itemize}
    Therefore, \eqref{eq:prox_f2_row} is obtained by Eqs.~\eqref{eq:appendix_prox_f2_row_setN_solution}, \eqref{eq:appendix_prox_f2_row_setU_solution}, and \eqref{eq:appendix_prox_f2_row_setC_solution}.

\bibliographystyle{IEEEbib-arxiv}
\bibliography{09_refs}

\end{document}